%% file: main.tex
\documentclass[acmlarge,screen,nonacm]{acmart}
\AtBeginDocument{%
  }

\usepackage{longtable}
\usepackage{tabularx}
\usepackage{array}
\usepackage{multirow}

\AtBeginEnvironment{tabular}{\setlength{\tabcolsep}{3.5pt}}
\AtBeginEnvironment{longtable}{\setlength{\tabcolsep}{3.5pt}}

\let\IMWUTOriginalTextColor\textcolor
\renewcommand{\textcolor}[2]{%
  \ifstrequal{#1}{red}{#2}{%
    \ifstrequal{#1}{blue}{#2}{\IMWUTOriginalTextColor{#1}{#2}}}%
}

\newcommand{\quoteit}[1]{\emph{``#1''}}
\definecolor{revisionzl}{RGB}{255,0,0}
\DeclareRobustCommand{\revision}[1]{\begingroup\color{red}#1\endgroup}
\DeclareRobustCommand{\revisionzl}[1]{\begingroup\color{revisionzl}#1\endgroup}

\renewcommand{\revision}[1]{#1}
\renewcommand{\revisionzl}[1]{#1}

\begin{document}

\title[When Ubiquitous Health Data Stops Helping]{When Ubiquitous Health Data Stops Helping: Patients' Experience of Breakdowns in the Data-Interpretation-Action Loop at the End of Life}


\author{Yibo Meng}
\authornote{Corresponding author.}
\orcid{0009-0009-0370-5456}
\email{yibom2@illinois.edu}
\affiliation{%
  \institution{Cornell University}
  \city{New York}
  \state{New York}
  \country{United States}
}

\author{Xinru Wang}
\orcid{0000-0002-0213-6425}
\email{xinru.wang@smart.mit.edu}
\affiliation{%
  \institution{Singapore-MIT Alliance for Research and Technology}
  \city{Singapore}
  \country{Singapore}
}

\author{Zhuoyan Li}
\orcid{0000-0001-7143-9262}
\email{li4178@purdue.edu}
\affiliation{%
  \institution{Purdue University}
  \city{West Lafayette}
  \state{Indiana}
  \country{United States}
}

\author{Chengxi Zang}
\orcid{0000-0002-8244-9551}
\email{chz4001@med.cornell.edu}
\affiliation{%
  \institution{Cornell University}
  \city{Ithaca}
  \state{New York}
  \country{United States}
}

\author{Zhuoran Lu}
\orcid{0000-0002-1079-2043}
\email{zhuoranlu34@gmail.com}
\affiliation{%
  \institution{Purdue University}
  \city{West Lafayette}
  \state{Indiana}
  \country{United States}
}

\author{Zhicong Lu}
\orcid{0000-0002-7761-6351}
\email{zlu6@gmu.edu}
\affiliation{%
  \institution{George Mason University}
  \city{Fairfax}
  \state{Virginia}
  \country{United States}
}

\renewcommand{\shortauthors}{Meng et al.}

\begin{abstract}
Ubiquitous computing systems in healthcare often assume that greater data visibility improves awareness, interpretation, and action. In end-of-life care, however, \revisionzl{this assumption warrants scrutiny}. \revisionzl{Through semi-structured interviews with 25 terminal-stage patients as primary informants and 8 clinicians as supplementary informants}, \revisionzl{we examine patients' descriptions of how they encountered, interpreted, acted on, and disengaged from state signals, trajectories, algorithmic outputs, and behavioral audits}. \revisionzl{Our analysis identifies three recurrent patterns in these descriptions}. First, \revisionzl{participants described repeated and unresolved sensemaking when visible data lacked explanatory pathways}. Second, \revisionzl{they described outputs they could understand but whose suggested actions did not fit their changing bodily capacities, care goals, or decision-making roles}. Third, \revisionzl{they described difficulty disengaging when data remained embedded across care environments}. Rather than rejecting data altogether, \revisionzl{patients called for less data exposure and a reorientation toward existential support}. \revisionzl{Based on this analysis}, we situate data exposure as a central interaction problem in end-of-life ubiquitous computing and derive design implications for interpretive stability, care-context alignment, and existential support.
\end{abstract}



\maketitle

\section{Introduction}
\input{sections/introduction}

\section{Related Work}
\input{sections/related_work}

\section{Method}
\input{sections/method}
\section{Results}
\input{sections/result_refactored}

\section{Discussion}
\input{sections/discussion}

\begin{acks}
The researchers thank all participants for their outstanding contributions to this paper. This is particularly true when considering the physical and psychological obstacles this group may face. All authors are sincerely grateful for their contributions to the research.
\end{acks}

\bibliographystyle{ACM-Reference-Format}
\bibliography{data_exp}

\appendix
\section{ Interview Guide}
\subsection*{Interview Guide: Patients}

\begin{table}[H]
\centering
\caption{Semi-Structured Interview Guide for Patients}
\label{appendix:interview_patients}
\small
\begin{tabular}{p{3cm}p{6cm}p{6cm}}
\toprule
\textbf{Theme} & \textbf{Core Question} & \textbf{Possible Probes} \\
\midrule

System
& What kinds of health-related computational systems do you use or encounter in your daily life?
& What specific devices or applications are they? In what situations do you usually use them? Can you describe a recent instance? \\

\addlinespace
Usage Mode
& How are these systems typically used?
& Do you actively check them, or encounter them through prompts or notifications? How often do you interact with them in a day? Are there fixed times? \\

\addlinespace
Information Presentation
& What kinds of information do these systems usually present to you?
& Which types of information do you see most often? Which appear occasionally? Under what conditions do they appear? \\

\addlinespace
Understanding
& When you see this information, how do you usually make sense of it?
& What is your first reaction? Do you relate it to previous experiences? Do you check it multiple times before forming a judgment? \\

\addlinespace
Action
& In what situations does this information influence your subsequent plans or actions?
& What did you do at that time? Did you act immediately or later? \\

\addlinespace
Action Not Taken
& In what situations does this information not lead to any action?
& Did you continue your original plan or observe for a while? Did you return to check the information later? \\

\addlinespace
Changes Over Time
& Has your way of using these systems changed over time?
& Has the frequency of checking changed? When did this change begin? \\

\addlinespace
Information Selection
& In your daily life, how do you decide which information to pay attention to?
& Are there types of information you focus on more? How do you handle the rest? \\

\addlinespace
Boundary
& Over time, have you developed your own ways or boundaries of using these systems?
& In what situations do you check them less? After reducing checking, do you return to them? Under what conditions? \\

\addlinespace
Open-ended
& Do you have any thoughts or perspectives about these systems or the data they provide?
& Is there a particular experience that stands out to you that you would like to describe? \\

\bottomrule
\end{tabular}
\end{table}

\subsection*{Interview Guide: Clinicians}

\begin{table}[H]
\centering
\caption{Semi-Structured Interview Guide for Clinicians}
\label{appendix:interview_clinicians}

\small
\begin{tabular}{p{3cm}p{6cm}p{6cm}}
\toprule
\textbf{Theme} & \textbf{Core Question} & \textbf{Possible Probes} \\
\midrule

System
& What kinds of computational systems related to patients do you encounter in your daily work?
& In what situations are these systems typically used? \\

\addlinespace
Usage Context
& How are these systems used within your actual workflow?
& Are they used routinely or only in specific situations? \\

\addlinespace
Understanding
& How do you usually interpret the data or outputs generated by these systems?
& Do you combine them with other information? Do you compare or further evaluate them? \\

\addlinespace
Decision Relationship
& In what situations do these systems influence your decisions?
& In what situations do they not enter your decision-making process? \\

\addlinespace
Action Process
& How is this information translated into actual handling or actions in practice?
& Does it directly correspond to certain actions, or require further judgment? \\

\addlinespace
Patient Understanding
& From your observation, how do patients typically encounter and interpret these systems?
& What kinds of information do they tend to focus on? \\

\addlinespace
Differences in Understanding
& In what situations does patients' understanding differ from your clinical judgment?
& What types of information tend to lead to such differences? \\

\addlinespace
Variation
& Does the role of these systems change across different patients or stages?
& In what ways do these changes manifest? \\

\addlinespace
Information Selection
& How do you decide which information to attend to in practice?
& Are there priorities among different types of information? \\

\addlinespace
System Position
& In your view, what role do these systems play in the overall care process?
& Does this role change across different situations? \\

\addlinespace
Open-ended
& Do you have any additional thoughts about the role of these systems in practice?
& Can you describe a representative case that illustrates your experience? \\

\bottomrule
\end{tabular}
\end{table}

\section{ Rationale for Excluding Family Caregivers from the Study}

\revisionzl{
Reasons for not including them in the care arrangements. The original sampling design did not include family caregivers as a third participant group for two reasons. First, our IRB-approved protocol limited recruitment to patients and clinicians because adding caregiver interviews could introduce additional privacy risks related to family relationships and increase participant burden in this particularly vulnerable population. Second, our research question focuses on patients' first-person experiences of encountering and living with ubiquitous health data. Interviewing patients therefore aligned with the scope of this study, while we explicitly acknowledge the absence of caregiver perspectives as a limitation and encourage future caregiver-inclusive research.

This decision should not be interpreted as suggesting that caregivers' perspectives are unimportant. Rather, it represents a methodological boundary of the present work. During the interviews, several patients described family members as participating in interpreting health data and making care decisions. Because caregivers were not interviewed directly, the study cannot cross-validate these patient accounts or provide a comprehensive multi-actor analysis of how patients, caregivers, and clinicians negotiate data interpretation, decision authority, and care practices. Future work should incorporate caregiver perspectives to better understand these collaborative processes.
}

\section{ Inclusion and exclusion criteria}
Given the extreme vulnerability of the target population, we established rigorous inclusion and exclusion criteria for the recruitment process. Specifically, for patients, the inclusion criteria were defined as follows: (1) a physician-confirmed diagnosis of a terminal illness with a limited life expectancy; (2) an initial physician assessment indicating sufficient physical and cognitive capacity to participate in the study, followed by successful completion of a subsequent cognitive assessment administered by the research team; and (3) the ability to fully comprehend the study's purpose, procedures, and potential implications, and to provide informed consent. Concurrently, the exclusion criteria included: (1) the presence of cognitive impairments or psychiatric disorders that compromise understanding and judgment; (2) a physical condition rendering participation in an interview medically inadvisable; (3) the presence of severe hearing or speech impairments; (4) being in a state of acute "emotional crisis" or extreme psychological instability; and (5) exhibiting strong resistance toward smart devices or related technologies such that the individual had not used or meaningfully encountered relevant digital health devices or data and therefore could not describe experiences pertinent to the study’s research question. This criterion did not exclude participants who had previously used such technologies but later reduced, discontinued, avoided, or refused their use; such experiences were considered directly relevant to our investigation of disengagement and refusal.

\section{Participant Vulnerability and Protection}

\revisionzl{Considering the vulnerability of participants, the interview protocol itself was also designed with this population's vulnerability in mind. As described in Section 3.2, the "system–information–understanding–action–boundary" sequence moved from concrete, low-burden questions about system use toward more interpretive and evaluative topics only after participants had established comfort with the conversation, rather than opening with questions about their condition or emotional state. For the same reason, the protocol deliberately avoided direct questions about "problems" or "negative experiences," which risked prompting defensive responses or unnecessary distress in a population already confronting significant physical and emotional burden; instead, such experiences were allowed to surface indirectly through concrete narrative questions. This design reflects a deliberate trade-off between eliciting rich experiential data and minimizing the emotional cost of participation for terminal-stage patients.}

\end{document}

%% file: sections/introduction.tex
Ubiquitous health infrastructures~\cite{bardram2005mobility,tentori2012pervasive}, from bedside monitors~\cite{drew2014insights,wong2015send} to wearables and mobile health applications~\cite{mamykina2008mahi,yasmin2020influence,meng2026living}, have extended sensing, visualization, and decision support~\cite{saleemi2023ubiquitous,yang2022review} far beyond isolated clinical encounters. From a human-centered perspective, \revisionzl{they are often framed around a recurring assumption that we later abstract as a ``data--interpretation--action'' loop}: richer and more continuous data should help patients, caregivers, and clinicians notice changes earlier, interpret conditions more accurately, and respond more effectively~\cite{li2010stage,epstein2015lived,mamykina2015adopting}. \revisionzl{This premise appears especially compelling in \textit{end-of-life care} during the late stage of a terminal illness}, \revisionzl{when care shifts from cure to comfort and focuses on symptom management and quality of life}~\cite{singer1999quality,teno2004family,lorenz2008evidence}. Here, conditions may change quickly, and care unfolds across different environments~\cite{sallnow2022report,temel2010early}---precisely the kind of setting where continuous sensing seems most valuable. For both patients~\cite{bhat2023we,miller2016partners} and clinicians~\cite{bardram2006awaremedia,basch2017overall}, then, continuous data visibility appears to offer reassurance, timely alerts, and better support for action when patients' conditions are fragile and the stakes of care are high.

\revisionzl{Realizing this promise requires continuous data to support interpretation and action in practice~\cite{andersen2011designing}. Prior work suggests that this process can break down in several ways. Health information can be fragmented, poorly framed, or too uncertain to support understanding}~\cite{senathirajah2014clinician,lamas2018advance,dillon2017provider,oechsle2019current,hebdon2023hospice,beil2019ethical}. \revisionzl{When monitoring involves self-tracking, continued participation can become burdensome and may eventually stop}~\cite{ancker2015you,epstein2016beyond,meng2026can,meng2026engagement}. \revisionzl{Systems may also cease to fit as users and situations change, a longstanding concern in ubiquitous computing}~\cite{schilit1994context,meng2026does}. \revisionzl{Together, these studies suggest that breakdowns in interpretation, engagement, and contextual fit extend beyond end-of-life care. In personal-informatics models, however, collection, reflection, action, and re-engagement can recur over time. This recurrence may provide later opportunities to revisit a misunderstanding, adjust a system, or resume tracking}~\cite{li2010stage,epstein2015lived,mamykina2015adopting}. \revisionzl{Research on mortality and technology shows that dying changes how people experience time and evaluate interactions}~\cite{massimi2011matters,kaptelinin2016making,sallnow2022report}. \revisionzl{At the end of life, a later opportunity to address a breakdown may not be available. A failed interaction may consume time, add burden, or reduce comfort without being corrected later. We therefore ask how patients experience these breakdowns when repair may not be possible and when failed interactions consume part of their remaining time.}




\revisionzl{This paper uses the data--interpretation--action loop to examine how these problems unfold at the end of life.} \revisionzl{We treat the loop as an analytical abstraction of recurring assumptions in prior personal-informatics research, not as a new process model.} \revisionzl{We ask:} \textit{How do terminal-stage patients encounter, interpret, and respond to health data made visible to them through ubiquitous computing systems in end-of-life care?} \revisionzl{We conducted semi-structured interviews with 25 patients as primary informants and 8 clinicians involved in their care as supplementary informants}. \revisionzl{We use four recurrent forms of health data---state signals, trajectories, algorithmic outputs, and behavioral audits---to organize participants' descriptions of data encounters}, \revisionzl{while treating these forms as participant-grounded empirical distinctions rather than a universal taxonomy}. \revisionzl{Through this lens, we identified three related ways in which encounters with health data became problematic. First, some visible data did not support an interpretation that patients could meaningfully use. Second, even when participants could interpret an output, the actions it implied no longer fit their bodily capacities, care priorities, or relationships with those involved in decision-making. Third, patients could not simply exit this data--interpretation--action loop, because exposure to data was often embedded in the technologies and routines of care. Taken together, these problems indicate that the value of health data at the end of life depended not only on what the data represented, but also on the technologies and care settings through which they were encountered, participants' relational and cultural circumstances, and the institutional resources available to support interpretation and response. Patients therefore did not reject health data altogether. Rather, they called for more selective data exposure and forms of support oriented toward comfort and personhood instead of continued bodily optimization.}

\revisionzl{Our contribution is to explain why familiar problems of interpretation, action, and disengagement take a distinct form under end-of-life conditions.} \revisionzl{We further develop the idea of \textit{temporal irreversibility}}: \revisionzl{later repair may improve what happens next}, \revisionzl{but it cannot recover time and attention already spent}, \revisionzl{and patients may have few opportunities to try again}. \revisionzl{From these findings, we derive design implications for interpretive stability, care-context alignment, and existential support.} \revisionzl{This analysis shifts the design question from how to maximize data visibility to how systems can support interpretation, feasible action, or refusal within patients' remaining time.}



%% file: sections/related_work.tex
\subsection{Ubiquitous Computing for End-of-Life Care}

\revisionzl{End-of-life data exposure spans both general-purpose clinical infrastructure and technologies developed for advanced illness or palliative care}. \revisionzl{In inpatient settings, bedside physiological monitors display real-time vital signs and alarms}~\cite{drew2014insights}. \revisionzl{Beyond hospital rooms, remote monitoring systems transmit patient-reported symptoms or physiological measurements to care teams}~\cite{mcgillion2018postoperative,basch2016symptom}. \revisionzl{Palliative telehealth supports home-based care and communication while also raising implementation, methodological, and ethical questions}~\cite{lundereng2023health,tarbi2025telehealth,krieckemans2025telehealth}. \revisionzl{Together, these technologies distribute data exposure across hospital rooms, homes, and mobile interfaces}.

\revisionzl{Many systems that expose end-of-life patients to data originate in broader hospital, chronic-disease, or wellness contexts}. \revisionzl{End-of-life patients' bodily conditions can change rapidly within a compressed and uncertain time horizon}, \revisionzl{while the goal of care may shift from cure toward symptom management, comfort, and quality of remaining life}~\cite{sallnow2022report,temel2010early}. Cognitive capacity may be diminished, routines of self-management interrupted, and decisions increasingly shared with family caregivers and clinicians under deep uncertainty~\cite{bhat2023we,miller2016partners}. \revisionzl{Health data can consequently acquire particular existential weight as patients encounter readings in relation to decline and remaining time}~\cite{massimi2009dying,kaziunas2018lived}. \revisionzl{Design assumptions from neighboring chronic-disease or wellness settings may not transfer unchanged to end-of-life care}; \revisionzl{how patients live with the data presented to them therefore warrants examination in its own right}.

\subsection{Health Data Exposure to Patients: Promise and Critique}

Within HCI and health informatics, patient-facing health data have long been framed as a route to empowerment through reflection. Personal informatics models posit a loop in which collection, integration, and reflection enable users to act on their own data~\cite{li2010stage,epstein2015lived,epstein2020mapping}, and a substantial body of work has built on this premise to support self-tracking and patient--provider collaboration~\cite{mamykina2008mahi,chung2016boundary,raj2019clinical,chen2026between}. The same premise underwrites clinical deployments of patient-reported outcome and continuous monitoring systems, where richer data are expected to translate into earlier detection and more effective response~\cite{basch2016symptom,basch2017overall}.

\revisionzl{Here, we use the ``data--interpretation--action loop'' as an analytical abstraction for a recurring assumption across this literature: patients encounter health data, construct an understanding of what those data mean, and use that understanding to decide whether and how to act. Our use of \textit{sensemaking} is grounded in \revisionzl{Dervin's formulation of how people confront a gap in a particular situation and construct a bridge that enables them to move forward}~\cite{dervin1992mind}, and in \revisionzl{Weick's conceptualization of sensemaking as an ongoing, socially situated process through which ambiguous circumstances are organized into a plausible interpretation that can serve as a basis for action}~\cite{weick1995sensemaking,weick2005organizing}. This perspective has also informed research on chronic-disease self-management, which distinguishes perceiving new health information, developing inferences that guide the selection of actions, and acting in response~\cite{mamykina2015adopting}. Accordingly, we use \textit{sensemaking} to denote the process through which patients mobilize available signals, prior experience, medical knowledge, and social resources \revisionzl{to bridge the gap between what displayed data show and what those data might mean in their particular situation}. We call the provisional understanding produced through this process \textit{interpretation}. A breakdown does not imply that patients understand nothing or take no action; \revisionzl{it occurs when available cues and resources do not enable patients to stabilize a workable interpretation or connect that interpretation to a feasible action}. We use these concepts as a sensitizing lens, rather than claiming a full application of either Dervin's methodology or Weick's organizational framework.}

A growing body of critical work, however, suggests that this loop does not reliably close. Patients abandon self-tracking tools when life circumstances shift or when tracking becomes burdensome~\cite{epstein2016beyond}; managing one's own data introduces invisible work and emotional labor that designs rarely acknowledge~\cite{ancker2015you,meng2026data}; the datafication of care reshapes relationships in ways that exceed the frame of individual self-management~\cite{kaziunas2017caring,kaziunas2018lived}; and informatics interventions intended to extend care often fail in the very populations they are meant to support~\cite{veinot2018good}.
\revisionzl{Personal-informatics and chronic-disease self-management research explains how people collect, integrate, interpret, and act on data}, \revisionzl{while critical self-tracking research shows how this work can become burdensome, relationally distributed, or abandoned when circumstances change}~\cite{li2010stage,epstein2015lived,epstein2016beyond,ancker2015you,kaziunas2017caring}. \revisionzl{Much of this literature nevertheless begins with tracking that patients initiate or manage.} \revisionzl{It offers less explanation of mixed data environments in which patient-facing applications coexist with clinician-facing infrastructure and exposure may continue even when a patient no longer chooses to inspect or act on the data.}

\revisionzl{Context-aware computing establishes that system behavior should respond to changing users and situations}~\cite{schilit1994context,weiser1999computer}. \revisionzl{This tradition explains why adaptation matters}, \revisionzl{but it does not by itself specify which aspects of an end-of-life situation determine whether a health-data output remains actionable}. \revisionzl{Our analysis therefore treats bodily capacity, care priorities, and decision relationships as empirically identified conditions of actionability, rather than presenting context adaptation or ``context drift'' as a new construct.}

\revisionzl{Thanatosensitive HCI and critical-data research foreground how mortality, temporality, identity, and relationships shape technology use}~\cite{massimi2009dying,massimi2011matters,kaptelinin2016making,kaptelinin2018technology,kaziunas2018lived}. \revisionzl{These traditions establish why end-of-life technologies cannot be understood only through efficiency or behavior change.} \revisionzl{What remains underexamined is how continuous exposure to clinical and patient-facing data repeatedly solicits interpretation and action, and why attempts to disengage from those demands may remain unstable.}

\subsection{Empirical Studies on Ubiquitous Computing for End-of-Life Patients}

\revisionzl{Research on digital technology in advanced illness and palliative care has followed three broad orientations. Outcome-oriented research evaluates whether symptom monitoring, web-mediated follow-up, home monitoring, or teleconsultation improves survival, symptom control, hospitalization, or quality of life}~\cite{basch2016symptom,basch2017overall,denis2017web,mooney2017automated,hoek2017effect,mao2024electronic}. \revisionzl{Design and implementation studies examine how these technologies are developed, integrated into care, and experienced by clinicians}~\cite{allsop2019paincheck,lundereng2023health,tarbi2025telehealth}. \revisionzl{Critical and ethical work examines the sociotechnical imaginaries and normative tensions surrounding palliative technologies}~\cite{ahmadpour2023hci,krieckemans2025telehealth}.

\revisionzl{Across these orientations, patients are rarely examined as \textit{interpretive subjects} who must determine what health data mean in their own situations. Outcome-oriented research emphasizes measured effects, design studies focus on development and workflow, and critical work foregrounds broader values and technological futures. Consequently, these literatures provide limited insight into how end-of-life patients interpret health data and decide whether or how to respond. Such questions cannot be inferred from outcome metrics, workflow descriptions, or interface evaluations alone, and overlooking them risks missing the legal, social, and ethical consequences of data infrastructures}~\cite{cohen2020ethical,mittelstadt2017ethics,hall2019moving}.

\revisionzl{The gap therefore lies not in tracking, context, mortality, or palliative technology individually, but at their intersection: how terminal-stage patients make sense of, act on, and attempt to disengage from health data embedded in care infrastructures. We examine this problem through the data--interpretation--action loop as an analytical abstraction rather than a new model, treating the four data forms as participant-grounded organizing categories. Our contribution is to show how unresolved interpretation, misaligned action, and unstable disengagement formed a distinctive configuration of end-of-life data encounters within a compressed time horizon where later repair could not be presumed.}

%% file: sections/method.tex
Studying how patients perceive their exposure to data from ubiquitous computing systems at the end of life poses a methodological challenge: the experiences are diffuse, unfold over time, and involve a population for whom research participation itself can be burdensome. We therefore conducted a qualitative study with 25 patients in the terminal stage of illness as our primary informants, supplemented by interviews with 8 clinicians involved in their care. This section describes our recruitment, data collection, analysis, and ethics reviews.

\subsection{Participants}

\revisionzl{To address our patient-centered research question, we recruited patients with terminal illnesses as the primary informants because the unit of analysis was patients' first-person experience of data exposure. We also recruited attending physicians as supplementary informants because many of the data encounters described by patients involved clinician-facing infrastructure, clinical interpretation, and hospital workflow. Clinician interviews were therefore used to contextualize and deepen patient-derived themes by clarifying relevant clinical and structural conditions; they were not treated as an independent dataset.We did not recruit caregivers; the reasons are provided in Appendix B.}

\begin{table}[H]
\label{appendix:demographics}

\centering
\caption{Demographic information of two types of participants: patients, whose interviews served as the primary empirical perspective, and clinicians involved in their care, whose interviews provided a supplementary perspective.}
\label{tab:appendix1}
\footnotesize
\renewcommand{\arraystretch}{0.92}
\begin{tabular}{llllllll}
\toprule
\textbf{ID} & \textbf{Role} & \textbf{Gender} & \textbf{Age} & \textbf{Region} & \textbf{Education} & \textbf{Diagnosis} & \textbf{Care Context} \\
\midrule
\multicolumn{8}{l}{\textit{Primary perspective: Patient participants}} \\
\midrule
P1  & Patient & Female & 63 & Urban & High School     & Advanced COPD                        & Hospital \\
P2  & Patient & Male   & 67 & Rural & Middle School   & Advanced COPD                        & Hospital \\
P3  & Patient & Male   & 45 & Rural & High School     & ALS with T2DM                        & Home     \\
P4  & Patient & Female & 44 & Urban & High School     & Advanced AIDS                        & Hospital \\
P5  & Patient & Female & 52 & Urban & Bachelor's      & Advanced AIDS                        & Hospital \\
P6  & Patient & Male   & 37 & Rural & High School     & ESRD (caused by T2DM)                & Home     \\
P7  & Patient & Female & 46 & Rural & High School     & ESRD                                 & Hospital \\
P8  & Patient & Male   & 76 & Urban & Bachelor's      & End-Stage Chronic Heart Failure      & Hospital \\
P9  & Patient & Male   & 41 & Urban & Doctoral        & ESRD                                 & Hospital \\
P10 & Patient & Female & 35 & Rural & High School     & Advanced AIDS                        & Hospital \\
P11 & Patient & Male   & 47 & Rural & Primary School  & ALS                                  & Hospital \\
P12 & Patient & Female & 55 & Rural & Bachelor's      & Advanced Lung Cancer                 & Hospital \\
P13 & Patient & Male   & 56 & Rural & Middle School   & Advanced Lung Cancer                 & Hospital \\
P14 & Patient & Male   & 37 & Urban & Bachelor's      & Advanced AIDS                        & Hospital \\
P15 & Patient & Male   & 33 & Rural & High School     & ESRD (caused by T2DM)                & Home     \\
P16 & Patient & Male   & 49 & Urban & Illiterate      & End-Stage Liver Cancer               & Hospital \\
P17 & Patient & Female & 71 & Rural & Middle School   & End-Stage Chronic Heart Failure      & Hospital \\
P18 & Patient & Female & 72 & Urban & Middle School   & ALS                                  & Hospital \\
P19 & Patient & Female & 56 & Rural & High School     & Advanced AIDS                        & Hospital \\
P20 & Patient & Female & 55 & Urban & Bachelor's      & ALS                                  & Hospital \\
P21 & Patient & Male   & 44 & Urban & Primary School  & ESRD (caused by T2DM)                & Home     \\
P22 & Patient & Female & 67 & Rural & High School     & End-Stage Chronic Heart Failure      & Hospital \\
P23 & Patient & Male   & 46 & Urban & High School     & End-Stage Liver Cancer               & Hospital \\
P24 & Patient & Male   & 25 & Rural & Bachelor's      & Multiple Sclerosis (MS)              & Hospital \\
P25 & Patient & Male   & 36 & Urban & High School     & End-Stage Liver Cancer               & Hospital \\
\midrule

\multicolumn{8}{l}{\textit{Supplementary perspective: Clinician participants}} \\

\midrule
P26 & Doctor  & Female & 44 & Urban & Doctoral        & N/A                                  & N/A      \\
P27 & Doctor  & Male   & 45 & Urban & Doctoral        & N/A                                  & N/A      \\
P28 & Doctor  & Female & 54 & Urban & Doctoral        & N/A                                  & N/A      \\
P29 & Doctor  & Male   & 39 & Urban & Doctoral        & N/A                                  & N/A      \\
P30 & Doctor  & Male   & 45 & Rural & Doctoral        & N/A                                  & N/A      \\
P31 & Doctor  & Male   & 46 & Urban & Doctoral        & N/A                                  & N/A      \\
P32 & Doctor  & Female & 54 & Urban & Doctoral        & N/A                                  & N/A      \\
P33 & Doctor  & Female & 52 & Urban & Doctoral        & N/A                                  & N/A      \\
\bottomrule
\end{tabular}
\end{table}

With IRB approval from the authors' institution, recruitment was conducted primarily within public hospitals using a combination of on-site recruitment and snowball sampling. The authors established connections with several public tertiary hospitals\footnote{Tertiary hospitals in China refer to institutions with the highest level of medical resources within a given region.} and initiated on-site recruitment in hospitals that supported the study. Participants were subsequently encouraged to refer other eligible individuals within the participating hospital network, following a snowball sampling strategy. To ensure diversity, we aimed to recruit participants with varied demographic characteristics, including gender, age, educational background, and illness status. In the context of end-of-life care, this further encompassed a range of illnesses and corresponding care approaches adopted by participants (e.g., home-based versus hospital-based care). Given the vulnerability of this target population, we established strict inclusion and exclusion criteria. See Appendix C for the criteria.

\revisionzl{Ultimately, we recruited 25 patients as primary informants and included 8 clinicians as supplementary informants.} Among the patient participants, \revisionzl{11 were female and 14 were male}, with ages ranging from 25 to 76. Among the 8 clinician participants, 4 were female and 4 were male, with ages ranging from 39 to 54. The sampled participants reflect diversity in educational backgrounds, ranging from illiterate to doctoral degree holders, and geographic origins, including both urban and rural regions. In the context of end-of-life care, patient participants also varied in medical conditions, including advanced lung cancer, end-stage renal disease, amyotrophic lateral sclerosis, end-stage chronic heart failure, end-stage liver cancer, advanced AIDS, advanced COPD, and multiple sclerosis, as well as care arrangements, spanning both home-based and hospital-based care, and prior experience with intelligent systems.See Table 1 for details.

\subsection{Data Collection}
We employed semi-structured interviews to understand the interaction process between users and intelligent systems, and users' perceptions, concerns, and expectations of intelligent systems for end-of-life care. Each interview lasted between 45 and 75 minutes.\footnote{Given participants' physical vulnerability, multiple breaks were often inserted during interviews (not counted toward total duration)} Interviews were conducted in-person and audio-recorded with participants' consent. Separate protocols were developed for the two participant groups (patients and clinicians), described below.

For patients, the interview structure progressively unfolded their everyday experiences with data from ubiquitous health systems, drawing on prior work on patient engagement with healthcare computational systems~\cite{tieu2015barriers,mcalearney2019patients}, sensemaking of system information~\cite{suija2022experiences,mamykina2015adopting,hohenstein2018supporting}, actions translated from that understanding~\cite{dai2026exploring,li2010stage,rooksby2014personal}, and boundaries patients draw around continued use~\cite{epstein2016beyond,satchell2009beyond,homewood2023self}. We consolidate these threads into a single experiential chain of ``system--information--understanding--action--boundary'' that structures each interview around the trajectory from information delivery to interpretation, action, and boundary-setting. \revisionzl{Here, we use \textit{usage boundaries} to refer not to fixed limits but to participants' evolving attempts to regulate data exposure and engagement, including reducing how often they checked data, selectively attending to particular data, avoiding displays or notifications, and deciding when to re-engage.} Interviews began by asking participants what computational systems they used and how, establishing a situational foundation, then moved to the information layer, focusing on how systems present information and how participants interpret it. Behavioral questions then explored under what conditions this information influenced or did not influence subsequent actions, capturing the structural relationship between information and behavior. Questions about "changes in use," "usage boundaries," and "information selection" guided participants to describe adaptive strategies developed over time, including reduced checking, selective attention, or active avoidance. The structure deliberately avoided directly asking about "problems" or "negative experiences"; instead, a continuous chain of questions enabled participants to naturally reveal their interpretive processes and behavioral patterns within concrete narratives, providing a basis for later identifying mechanisms such as interpretive loops, breakdowns of action, and strategic ignoring. Finally, participants shared their understandings, expectations, and any other topics regarding computational systems. The detailed protocol appears in Table~\ref{appendix:interview_patients} in the Appendix A.

For clinicians, the interview followed the same overall logic as for patients, but shifted focus to system use and interpretation in practice and relationships among multiple actors. Interviews began with the patient-related systems clinicians encounter daily and how these are used, situating systems within clinical workflows, then turned to how clinicians interpret system-generated data and outputs and how this information enters their decision-making and treatment processes. Questions on "decision relationships" and "action processes" then examined when systems did or did not influence clinical decisions, revealing the boundary between system outputs and professional judgment, while clinicians also described, from their observations, how patients engage with and interpret these systems and where discrepancies arise. Finally, questions on "usage differences," "information selection," and "system positioning" explored how system roles shift across contexts and patient conditions. This structure enabled \revisionzl{clinicians to discuss more than} system functionality, including interpretive processes in practice and cross-actor interactions, providing multi-perspective evidence for analyzing tensions among patients, clinicians, and computational systems. The detailed protocol appears in Table~\ref{appendix:interview_clinicians} in the Appendix A.

\subsection{Ethics}
This study received approval from the institutional ethics committee and strictly followed the principles of respect, beneficence, justice, and non-maleficence.

Prior to formal interviews, participants underwent multiple rounds of screening and confirmation. First, attending physicians assessed their level of consciousness and physical stability. After that, researchers conducted capacity assessments based on adapted evaluation principles, confirming through dialogue whether participants possessed the ability to understand the study content, engage in reasoning and judgment, and express autonomous choices. Only after confirming full understanding did the informed consent process proceed. The study content was explained by accessible language, with core information repeated at least three times and comprehension check through paraphrase. All participants signed written informed consent forms and were explicitly informed of their rights to withdraw at any time, skip questions, or request removal of their data. Finally, participants were clearly informed that the study would not provide direct medical benefits, but might offer opportunities for emotional expression and meaning-making. Considering that monetary compensation may be inappropriate or ineffective for participants in this context, all participants received a bouquet of flowers after the study\footnote{The flowers were selected according to participants' allergies and preferences, with consent from both the participant and physician.} as a symbolic gesture of appreciation. All clinical staff assisting the study were compensated 15 USD.

For participants' emotional safety, all interviewers received training in palliative care communication and psychological support. During interviews, participants' conditions were continuously monitored. A professional psychological counselor was on standby to provide support and terminate the process if any distress arose. In addition, a psychological support hotline was offered to address potentially delayed emotional responses.

Audio recordings were securely encrypted and stored, and original audio files were deleted after transcription. Textual data were anonymized, with all identifiable information removed, and access was restricted to the core research team. Potentially identifiable details were further obscured in reported results.

\revisionzl{Furthermore, considering the vulnerability of the participants, our interview design sought to avoid causing negative impacts on them (as shown in Appendix D).}

\subsection{Data Analysis}

\revisionzl{
We analyzed the interview data using reflexive thematic analysis~\cite{braun2019reflecting,braun2021one}, understanding themes as interpretive claims actively constructed and generated by the researchers within the data. Three authors were involved throughout the analytic process.

All recordings were transcribed verbatim in Chinese and translated into English for analysis. The original transcripts were retained and repeatedly consulted throughout the analysis to ensure that semantic nuance was not diminished by the translation process. Each researcher then read through the full set of materials independently. During familiarization, we paid particular attention to how participants described their bodily conditions, their processes of meaning making, and their interactions with digital systems, annotating initial impressions (e.g., ``confusion,'' ``dependence,'' ``avoidance'') to surface analytic leads. \revisionzl{Throughout the analysis, patient transcripts served as the primary evidence for experiential themes, while clinician transcripts were examined as supplementary material that contextualized and deepened those themes by clarifying clinical interpretation, infrastructure, workflow, and resource constraints.}

The three researchers each engaged with the full dataset and developed initial codes and analytic interpretations. These individual readings were subsequently discussed collaboratively, with differences treated as opportunities to examine alternative interpretations and refine the developing analysis rather than as discrepancies requiring agreement or resolution~\cite{braun2021can}. Illustrative initial codes included passive monitoring (continuously checking data without understanding it), active interpretation (making self judgments from data), and emotional triggering (affective reactions to data).

Codes were then collaboratively clustered into candidate themes and reviewed against the full dataset for internal coherence and external distinctiveness. Themes were defined in terms of the interpretive claim each made about participants' experiences, rather than as surface descriptions of their constituent codes; final theme names were chosen to convey those claims themselves (e.g., ``data as cognitive burden,'' ``the interpretive impulse and unresolvable loops,'' ``from understanding data to escaping data'').

To support analytic rigor, themes were systematically traced back to the original transcripts to ensure they were well grounded in the data, and the research team engaged in ongoing reflexive discussion to surface and examine our own assumptions and positionality.}

\revisionzl{\revisionzl{Rather than treating saturation as a universal criterion for sample adequacy}~\cite{braun2021saturate}, the research team judged that the accumulated patient interviews provided sufficient experiential depth, variation, and interpretive richness to support the explanatory analysis presented in this paper.}

%% file: sections/result_refactored.tex
To answer our core research question, we organize the findings in three stages. First, we characterize end-of-life care as a pervasive health-data environment and identify four recurrent data types. Second, \revisionzl{we examine the difficulties participants described in interpreting and acting on these data, and analyze these difficulties in terms of sensemaking and context management}. Third, \revisionzl{we examine why participants found it difficult to exit this data environment fully and how they characterized this unstable state of partial disengagement}. \revisionzl{Unless otherwise indicated, the term ``participants'' in this section refers to patient participants; clinician interview material is explicitly identified and used to contextualize the patient-derived findings.}

\begin{figure*}[!htbp]
  \includegraphics[width=\textwidth]{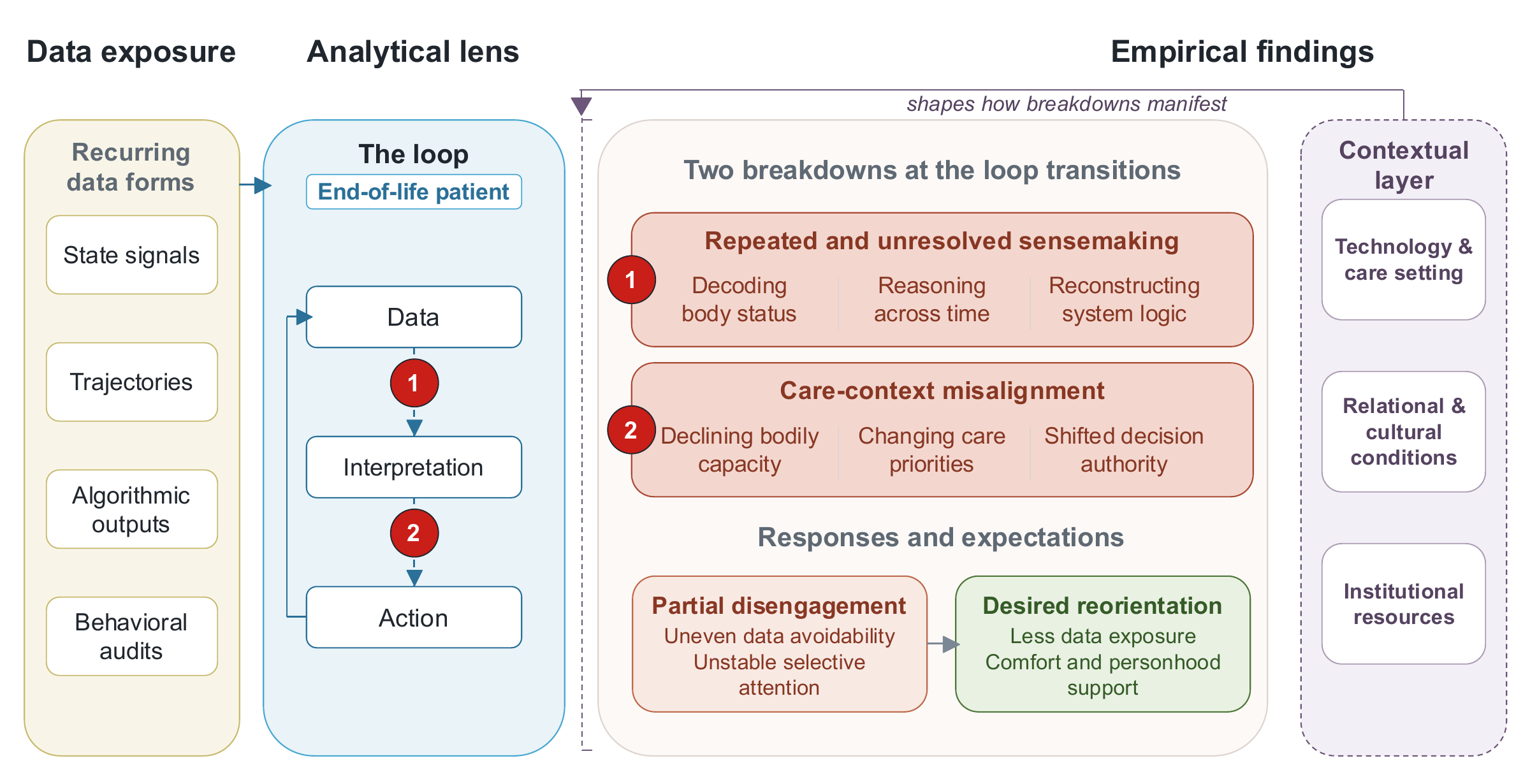}
  \caption{\revisionzl{Framework overview.} \revisionzl{The left panel summarizes four recurrent forms of health data encountered by patient participants, and the blue panel presents the data--interpretation--action loop used as the analytical lens.} \revisionzl{Numbered red markers at the data-to-interpretation and interpretation-to-action transitions connect the lens to two recurrent breakdown patterns identified in participant interviews: repeated and unresolved sensemaking and care-context misalignment.} \revisionzl{The lower row distinguishes patients' partial disengagement from their desired reorientation toward less data exposure and support for comfort and personhood.} \revisionzl{Within the empirical-findings region, the contextual layer summarizes technological and care-setting conditions, relational and cultural conditions, and institutional resources.} \revisionzl{Its backward connector points to the boundary between the analytical lens and empirical findings, indicating our interpretation that these conditions influenced how the reported patterns appeared rather than constituting another stage of the loop.}}
  \label{fig:workflow}
\end{figure*}

\subsection{Living with Data: Health Information Contexts in End-of-Life Care}
\label{results:data-environment}

\revisionzl{Participants described health data not merely as objects they requested on demand (e.g., by checking an Apple Watch), but as an information environment embedded in their everyday lives}. \revisionzl{They described how data streams from medical sensing devices and health-related applications were persistently generated, updated, and encountered}. \revisionzl{22 patient participants explicitly reported that health data were involved in their lives to varying degrees.} \revisionzl{Based on recurring differences in the interpretive and practical work participants described, we use four recurrent forms to organize the findings}: \textit{state signals, trajectories, algorithmic outputs, and behavioral audits}. \revisionzl{These are participant-grounded organizing categories rather than a proposed context-independent taxonomy}. \revisionzl{They resonate with prior models of personal-informatics stages and prior analyses of sensemaking in chronic-disease self-management}~\cite{li2010stage,epstein2015lived,mamykina2015adopting}, \revisionzl{while allowing us to trace how different forms of data acquired specific meanings and consequences in end-of-life care}.

\revisionzl{The remaining 3 participants, P22, P23, and P25, mentioned experiences of exposure to health data but described its impact on their lives as relatively limited, a pattern related to the nature of their illness, physicians' assessments, and personal mindset. For example, P22 said, \quoteit{My condition is stable, there's no more treatment, and I won't die suddenly, that's what the doctor said, so I don't have that bedside machine, and aside from my medical record I rarely encounter data, so I'm not too anxious about it.} P25, meanwhile, said, \quoteit{Ah, it's not a big deal, I can't be bothered to look, there's only so much time left, why worry about all this stuff.}}


\noindent \textbf{State Signals.}
The first category consists of raw physiological measurements (e.g., heart rate, oxygen saturation, blood glucose, blood pressure, and respiratory rate). In end-of-life care, these signals are high-frequency, near real-time, and primarily descriptive. \revisionzl{Participants described receiving these signals without interface explanations and having to interpret whether changes were concerning or actionable, sometimes together with family members.} \revisionzl{Participants characterized exposure in end-of-life care as shaped less by user-initiated tracking than by passive encounters within the care environment}. As P11 explained, \emph{``that monitor by the bed is always on,''} and values are \emph{``very easy to see''} in daily life. \revisionzl{Participants reported encountering this type primarily through bedside monitors or wearable devices}. \revisionzl{Hospital-based participants described this exposure as difficult to refuse and reported that the data remained within their field of view}. \revisionzl{For example, P28, a clinician,}  \revisionzl{explained, \quoteit{The data at the bedside is mainly for the doctors to look at. Sometimes patients ask me about it, and I just tell them not to worry about it.}}


\noindent \textbf{Trajectories.} The second type of data includes trend curves, fluctuation ranges, and prompts that orient patients toward change over time. Here the issue is no longer only what the body is doing now, but where it seems to be going. As P6 explained, \quoteit{there's a full-day glucose curve. I check it every day---morning, afternoon, and evening. Sometimes I pull up previous days to compare, to see if there's a downward or upward trend.} \revisionzl{P15 similarly described an app that presented labels such as} \quoteit{large fluctuation} \revisionzl{alongside the graph; he explained}, \quoteit{I usually look at the curve first, and then read those prompts.} \revisionzl{Participants described these data as organizing experience temporally}. \revisionzl{What distinguished trajectory data was the directional reading participants described}: \revisionzl{trends appeared not only as patterns of variation but as indications of where the body was heading overall}. P6 articulated this directly: \quoteit{I look at the trend to see how much time I have left. When the curve goes down, I feel like there isn't much left.} \revisionzl{Participants thus described trends as carrying an additional meaning about remaining time beyond what the curves were designed to display}. \textcolor{red}{\revisionzl{Participants reported encountering this type through bedside monitoring devices and mobile applications}. \revisionzl{They generally described accessing applications voluntarily, } \revisionzl{while P14 reported doing so at a physician's request}.}

\noindent \textbf{Algorithmic Outputs.} The third type includes algorithmically inferred outputs such as risk levels, abnormal-state labels, and recommendation snippets. Participants often \revisionzl{described encountering} these data not as raw inputs to interpret, but as conclusions that had already been drawn for them. P18 said, \quoteit{Sometimes the app pushes suggestions like `reduce activity' or `pay attention to rest.' You see them directly in the notification bar or when you open the app.} Likewise, P9 explained, \quoteit{It will directly tell you what state you are in, like `needs attention.' It's shown at the very top of the homepage, so you don't have to look into the detailed data yourself.} What makes this type distinct is that it carries a stronger evaluative force: it does not just show a bodily state, but names it, classifies it, and implicitly tells the patient that something should now be noticed or acted upon. \textcolor{red}{\revisionzl{Participants reported receiving this type through mobile applications and wearable devices}.}

\noindent \textbf{Behavioral Audits.} \revisionzl{The fourth category consists of records that allow patients to review past health-related behaviors, including activity logs, dietary or medication records, and other completed or missed care-related actions}. Unlike the other three types of data, behavioral audits do not describe what the body is doing now, but rather what was done to or for the body. P13 noted, \quoteit{There's a section for recording diet and activity---it uses location to track how many steps I've taken. I usually check it at the end of the day; it gives a summary of the whole day.} \revisionzl{Participants' examples centered largely on basic activities such as diet and step counts; we therefore interpret these audits less as records of performance than as records of remaining capacity and whether basic care tasks had been completed}. Behavioral audits can also arise through interaction with other data types while remaining distinct from them: \revisionzl{in P5's interview, a medication reminder constituted an algorithmic output}, \revisionzl{while the subsequent record formed an audit that she could check for overlooked care tasks}. \textcolor{red}{\revisionzl{Participants reported encountering behavioral audits through mobile applications, wearable devices, and paper or electronic medical records}.} \revisionzl{It is worth noting that behavioral audit data were reported by most participants as having relatively little impact on their lives. A representative response was P4: \quoteit{Honestly, I think at this stage, this kind of summary, record, something meant to be responsible for or guide future matters, it no longer has any value, so even though I know this record exists, I really don't care about it anymore.}}

Having outlined these four data types, we now use them to guide the rest of the results. These categories are not meant to classify interfaces or system outputs. Instead, they capture the interpretive and practical work \revisionzl{participants described in relation to health data} in end-of-life care. This matters because the problems we identify next do not emerge in the same way across all four types. \revision{\revisionzl{We first discuss the difficulties participants described in interpreting data and how these difficulties varied across data types}.}

\revisionzl{\revisionzl{Participants were recruited across two care contexts, hospital-based care (N=21) and home-based care (N=4), and most reported data exposure in the hospital setting}. \revisionzl{Hospital-based participants reported encountering bedside monitoring devices, paper or electronic medical records, mobile applications, and wearable devices}, \revisionzl{whereas home-based participants did not report exposure to bedside monitoring devices}. \revisionzl{Hospital-based participants described bedside-monitor data as especially difficult to avoid}, \revisionzl{whereas participants described mobile applications, continuous glucose monitoring (CGM) devices, and other wearables as allowing greater scope for partial disengagement}. \revisionzl{In addition, all participants reported use of paper or electronic medical records, and both were widely regarded across both groups as relatively easy to avoid and as having no significant impact.} \revisionzl{Given the imbalance between the two groups, these interviews do not support a balanced comparison of settings, but our analysis suggests different possibilities for disengagement across the settings, as summarized in Tables}~\ref{tab:hospital_data_sources} and~\ref{tab:home_data_sources}.}

\begin{table}[t]
\centering

\caption{\revisionzl{Participant-reported data sources and modes of exposure in hospital-based care; intended-user classifications are analytical}.}
\label{tab:hospital_data_sources}
\scriptsize
\setlength{\tabcolsep}{4pt}
\begin{tabular}{p{1.8cm}p{2.3cm}p{2.6cm}p{1.8cm}p{2.2cm}}
\toprule
\textbf{Data Source or Device} & \textbf{Participants} & \textbf{Data Encountered} & \textbf{Intended Audience} & \textbf{Mode of Patient Exposure} \\
\midrule
Bedside physiological monitoring devices & P1, P2, P4, P5, P7, P8, P9, P10, P11, P12, P13, P14, P16, P18, P19, P20, P24 & Heart rate, oxygen saturation, blood pressure, respiratory rate, ECG waveforms, and trends & \revisionzl{Primarily clinicians, with patients as incidental or secondary viewers} & Continuously visible in the room, often without active patient initiation \\
\addlinespace
Mobile health applications & P2, P4, P5, P7, P8, P9, P10, P11, P12, P13, P14, P16, P18, P19, P20, P24 & Trend graphs, abnormality labels, reminders, risk classifications, recommendation snippets & Patients and, in some cases, family caregivers & Patient-facing notifications or information accessed via phones \\
\addlinespace
Wearable devices & P7, P14 & Physiological readings, activity records, alerts, and trends & Primarily patients & Continuously worn and periodically checked \\
\bottomrule
\end{tabular}
\end{table}

\begin{table}[t]
\centering

\caption{\revisionzl{Participant-reported data sources and modes of exposure in home-based care; intended-user classifications are analytical}.}
\label{tab:home_data_sources}
\scriptsize
\setlength{\tabcolsep}{4pt}
\begin{tabular}{p{1.8cm}p{2.3cm}p{2.6cm}p{1.8cm}p{2.2cm}}
\toprule
\textbf{Data Source or Device} & \textbf{Participants} & \textbf{Data Encountered} & \textbf{Intended Audience} & \textbf{Mode of Patient Exposure} \\
\midrule
Continuous glucose monitors & P3, P6 & Real-time glucose values, glucose curves, fluctuation trends, and alerts & Primarily patients and caregivers, with clinicians \revisionzl{as secondary viewers when data are shared} & Continuously available through sensors and linked apps \\
\addlinespace
Mobile health applications & P3, P6, P15, P21 & Glucose trajectories, health scores, reminders, behavioral records, abnormality labels, recommendations & Primarily patients and family caregivers & Accessed voluntarily, with some push notifications \\
\addlinespace
Wearable devices & P15 & Activity records, physiological measurements, and prompts & Primarily patients & Worn in everyday domestic settings \\
\bottomrule
\end{tabular}
\end{table}

\subsection{When Data Leave Questions Open: Repeated and Unresolved Sensemaking Across Data Types}
\label{results:unresolved-sensemaking}
\revisionzl{22 patient participants described repeatedly checking the same data and continually trying to understand what it meant, but these attempts often failed to produce a stable interpretation. P22, P23, and P25, who described limited data exposure and impact, were exceptions to this pattern. Across different types of data, many participants described trying to guess what the data might mean. Several participants recalled that when they saw a number drop, a curve change, or a label appear, they would typically try to guess the reason.} \revisionzl{Across these interviews, we interpret participants' reconstructions as unreliable links between observations and explanations when appropriate context was absent}. We describe this activity as \textbf{repeated and unresolved sensemaking}: patients' attempts to understand the meaning of data in the absence of an explanatory pathway.

\revision{\revisionzl{Based on the interview material, this pattern reflected a persistent gap between what systems displayed and what they explained}. \revisionzl{Participants described relying on fragmented medical knowledge and personal memory to bridge this gap, yet recalled that repeated sensemaking rarely converged on a clear answer}. \revisionzl{Later data often failed to resolve their existing doubts}. \revisionzl{Below, we examine participants' experiences with three types of data before discussing why they described this form of sensemaking less often for behavioral audits}.}

\noindent \textbf{Decoding the body through state signals.} \revisionzl{For raw state signals such as oxygen saturation or ECG readings, participants described trying to understand what might be happening in the body to explain a given value}. \revision{\revisionzl{Across participant interviews, we identify two characteristics that made these signals difficult to interpret conclusively}.} First, a single value carried limited meaning without medical knowledge. Second, \revisionzl{continuously changing values required temporal context for interpretation}. Participants \revisionzl{commonly described lacking} both.

\revisionzl{P7 illustrated this pattern}. She described being drawn in by a change in her state signals: \quoteit{Sometimes when my oxygen level changes from 96 to 94, I stare at it for a while.} \revision{\revisionzl{P7 recalled that} \revisionzl{waiting did not bring understanding; when the number rose again, her confusion resurfaced:}} \quoteit{But after a while it goes back up again, and then I start wondering again whether I saw it wrong just now.} \revisionzl{She worked backward toward a cause, asking whether} \quoteit{it's because I just moved a bit, or whether something is wrong.} \revisionzl{The displayed signal did not help her choose between the two explanations}.

One \revisionzl{condition participants associated with} this gap was limited access to medical context. P1 described her inability to make sense of the bedside monitor: \quoteit{I'm old, and when you look at the ECG beside the bed, I simply cannot understand what it means.} \revisionzl{P1 said that} seeking help from clinicians did not close the gap, but relocated it: \quoteit{if I ask the doctors, they do not explain it, and if I ask too much, they get annoyed.} \revisionzl{P26, a physician, contextualized this gap by identifying both systemic limits and a further constraint in the data itself:} \quoteit{I admit that in many cases I really did not explain these data clearly enough to patients. Ideally, I should have done so. But, as you know, given the medical conditions in our country, it is already difficult enough just to try to treat patients properly. I truly do not have time to explain all of this. And, more importantly, many data---such as an occasional abnormal fluctuation in an ECG---simply do not mean very much in themselves. Even if you asked me to explain them, I could not really explain them.} \revisionzl{We interpret what P26 described as evidence that the missing explanatory pathway arose through the interaction of absent interface explanation, limited clinical time, and signals that sometimes lacked stable clinical meaning, rather than from patient literacy or data presentation alone.}

\revisionzl{A second condition participants described was limited analytic support}; participants \revisionzl{described falling} back on fragmented personal memory. P1 recounted, \quoteit{I remember that the last time something abnormal happened with my heart and lungs, that ECG line was shaking a lot, so I kept watching whether that line changed.} Similarly, 12 patient participants, including P1, P6, P7, and P18, used expressions such as \quoteit{roughly, I may remember} (P6), \quoteit{seems there is a change} (P7), and \quoteit{that looks similar to something in the past} (P18), \revisionzl{suggesting a common reported approach of using memory as a reference}. \revisionzl{These comparisons generated further guesswork rather than closing it}.

\noindent\textbf{Reasoning \revisionzl{across time} through trajectories.} With trend visualizations, the object of sensemaking expanded along the temporal axis. \revision{\revisionzl{Several participants described using trends to compare data across time and reflect on what might have happened in the past or what might happen next}.}

\revisionzl{P15 provided an example of backward-looking sensemaking}: the trend \revisionzl{prompted a question that he lacked} the resources to resolve. \quoteit{Sometimes it says the fluctuations are large, and then I start thinking about what period of time something went wrong in ... but in fact I still cannot find a definite reason. Because the curve is always fluctuating. Still, I have noticed that some factors may have some related effects.} \revisionzl{Participants described this sensemaking across periods ranging from minutes (P12, P9) to days (P2, P18), and beyond}. By contrast, P6 illustrated the forward-looking side of this process, reading a descending curve as a question about trajectory: \quoteit{If the curve is going downward, I keep looking at it, to see whether it is still dropping. Sometimes I think about whether it will get worse next, but there is nothing I can do to change it.} \revision{\revisionzl{Several participants described inferring future developments from trends}, \revisionzl{although the data were often insufficient to support a clear judgment}.}

\revisionzl{Participants described this form of sensemaking as particularly burdensome when trend data were aggregated into composite scores that seemed to speak to overall condition}. P21 described how far this reasoning could extend: \quoteit{I often check my overall health score...especially when I see the curve continuing downward. I become anxious that maybe something throughout this whole period of life has gone wrong. It feels as if that curve is not going downward, but as if I myself am gradually approaching death.} \revisionzl{Here, sensemaking extended beyond the temporal context of the displayed data to encompass the course of a life}.

\noindent\textbf{Reconstructing \revisionzl{system logic from} algorithmic outputs.} \revision{\revisionzl{When systems provided semantically richer outputs, such as judgments or suggestions, several participants described trying to understand why the system had produced them}. \revisionzl{They typically tried to infer what information the system had used to reach each conclusion}. \revisionzl{Our analysis suggests that sensemaking was especially pronounced when the system's reasoning was not visible}.}

P16 described this pattern: \revisionzl{after the app told him that he was in an} \quoteit{abnormal state,} he recalled \quoteit{wondering whether it is because of some particular indicator, what caused it}, but said, \quoteit{I do not know which one.} \revisionzl{The conclusion arrived without the indicators that supported it}, \revisionzl{leaving him to guess} at them. \revisionzl{P19 described a more consequential version, in which absent explanations carried perceived clinical and social stakes}: \quoteit{The app is always telling me that one of my indicators is abnormal...At the beginning, every time the app said something was abnormal, I would feel like maybe I was about to die. Later, I gradually realized that many abnormalities are not actually that important...I can press this button to call a doctor, but I am afraid that if I call the doctor over and it turns out to be nothing serious, the doctor will be unhappy. But I also do not know which judgments indicate real danger.} \revision{\revisionzl{P19 described gradually drawing on personal experience to judge which alerts warranted attention}.} \revisionzl{P19 indicated that, without} access to the system's reasoning, she could not confidently distinguish which alerts warranted action. \revisionzl{This uncertainty was compounded by a perceived social cost that she associated with acting on them}.

\revisionzl{P27, a physician, also framed this as a systemic issue.} He noted that doctors often tell patients \quoteit{not to worry too much} because \quoteit{if something is truly dangerous, doctors can also see it in real time,} and acknowledged that \quoteit{the problem patients report is real.} \revisionzl{P27 identified} two sides. On the application side, \quoteit{many apps do not distinguish urgency or severity, and do not consider how these data and labels will be understood by users.} On the clinical side, oversight was not always a reliable backstop: \quoteit{there are indeed still some unexpected situations in which what doctors see may lag behind what patients experience.} \revisionzl{We therefore interpret the missing explanatory pathway described here as involving both a design problem and a limit of clinical oversight}.

\revision{Overall, \revisionzl{participants described a similar pattern across data types: they tried to infer meaning or causes from displayed data}, \revisionzl{but most reported that these efforts did not yield stable interpretations}. \revisionzl{Participants generally described systems as presenting results without the contextual information needed to understand them}, \revisionzl{leaving participants to infer their meaning and underlying rationale}. \revisionzl{Taken together, these findings suggest that current systems more readily support data presentation than stable interpretation}.}

\subsection{When Action Cannot Follow: Misalignment in Context Management}
\label{results:care-context-misalignment}

\revision{\revisionzl{Even when participants believed they understood the data, they described difficulty acting on it or found that the suggested actions no longer applied to their situations}.} \revisionzl{We interpret this breakdown not simply as a failure of recommendation quality, but as a misalignment in \textit{context management}}. \revisionzl{Participants described wearables, bedside monitors, and mobile health applications as continuing to produce suggestions, notifications, and behavioral prescriptions}. \revisionzl{The issue participants described was not necessarily the correctness of these outputs, but their fit with the patient's lived situation}. \revisionzl{Systems were described as continuing to address patients as if bodily condition, care priorities, and decision-making roles remained stable, even when participants reported that these had shifted in end-of-life care}.

\revisionzl{The need for ubiquitous systems to adapt to changing users and environments is a longstanding context-awareness problem}~\cite{schilit1994context,weiser1999computer}. \revisionzl{We therefore use \textit{context drift} descriptively}, \revisionzl{rather than as a new theoretical construct, to name the growing mismatch participants described between system assumptions and their lived situations}. \revisionzl{End-of-life care made this mismatch especially consequential because bodily capacity, care priorities, and decision relationships could change together within a compressed time horizon}. \revisionzl{Across the interviews, participants described three manifestations}: \revisionzl{the bodily capacity assumed by the system no longer matched their current condition}; \revisionzl{the temporal horizon optimized by the system no longer matched their priorities}; \revisionzl{and the person addressed by the system was no longer always the person they expected to act}. \revisionzl{The following sections trace these manifestations and the consequences participants described}.

\noindent \textbf{When the body can no longer act.} \revisionzl{A first manifestation participants described concerned a mismatch between the body assumed by the system and participants' lived bodily capacities}. \revisionzl{Participants described systems that continued to generate suggestions as if they were still able to walk, exercise, or adjust daily routines, and some reported that the recommended actions had become physically impossible as their capacities deteriorated}.

P11, who was bedridden, described how activity reminders no longer functioned as actionable guidance: \quoteit{Sometimes it tells me that I have too little activity and that I should walk more. But now I even have difficulty turning over in bed by myself, let alone walking around. In a sense this is actually easy to understand, because I really cannot do it, so there is nothing much to struggle about. But I really want to turn off this reminder, because it keeps telling me I have failed to complete the task. It feels as if it is constantly reminding me that I can no longer do anything.} \revisionzl{For P11, the breakdown involved not only being directed toward an action he could no longer take, but also experiencing repeated notifications as a recurring reminder of lost capacity}.

\revisionzl{Some participants described attempting to repair this mismatch through more adaptive tools such as AI-powered mobile conversational agents}. \revisionzl{They reported that these systems sometimes appeared to tailor recommendations once participants explicitly stated their condition; as they described it, adaptation depended on repeatedly re-establishing context}. \revisionzl{P24 described having to restate that exercise was no longer possible before his phone provided more appropriate suggestions}: \quoteit{Every time I report to my phone, I have to tell it that I can no longer exercise...it helps a bit, but it is exhausting, because every time before starting the conversation I have to explain it again.} \revisionzl{We interpret this as shifting context-management work} onto the patient. \revisionzl{P24 said that the system did not retain his changed capacity and required him to declare it repeatedly}.

\noindent \textbf{When goals no longer apply.} \revisionzl{A second manifestation participants described concerned actions they described as physically possible but no longer aligned with what they considered meaningful}. \revisionzl{We interpret this as a mismatch between a recommendation and the time horizon participants described living within}.

\revisionzl{Patient participants and P30 described the systems discussed in their interviews as embodying a long-term health-management logic}. \revisionzl{They described systems as assuming a future in which present self-management would pay off later, a temporal logic patient participants no longer considered relevant in end-of-life care}. P30 described this mismatch clearly: dietary intervention systems \quoteit{always try to intervene in life from every angle,} but for \quoteit{participants whose expected survival is only one year, making them worry about what might happen ten years later---don't you think that is ridiculous?} \revisionzl{P30 further described patients as lacking support for distinguishing which goals still mattered}: \quoteit{They find it very hard to distinguish between factors that determine whether they survive now, and those that affect ten years or even twenty years into the future.} \revisionzl{P30 thereby suggests that systems presented these different temporal stakes within the same stream of recommendations}.

\revisionzl{P10 described the same temporal reprioritization from a patient's perspective:} \quoteit{If it is the kind of long-term influence, like what may happen later on, I do not really think about that anymore. What matters more is whether it will have an obvious effect right now.} \revisionzl{Together, P10's focus on immediate effects and P6's reading of trajectories in relation to perceived remaining time indicate that immediacy and perceived remaining time formed part of how these participants judged whether data were relevant.}

\revisionzl{Participants often described this mismatch in terms of fatigue rather than explicit disagreement}. P3 put it bluntly: continuous glucose monitoring \quoteit{makes me feel exhausted.} P14 described a more constrained version, \revisionzl{reporting that he} \quoteit{already knew} that he \quoteit{no longer needed to pay attention to blood pressure issues,} \revisionzl{yet that the watch continued to remind him every day and that he kept wearing it because} \quoteit{the doctors asked me to record blood pressure, and they think these data are useful.} \revisionzl{P14 reported that the data stream was no longer relevant to him but remained useful to his doctors, indicating that its purpose had shifted toward other stakeholders while the burden of exposure remained with him.}

\noindent \textbf{\revisionzl{When control was described as having moved elsewhere.}} \revisionzl{A third manifestation participants described concerned data that still pointed to an action that mattered and could be taken, while the relevant decision was described as having shifted to a family member or physician.} \revisionzl{This finding concerns how patients experienced a changed decision relationship, rather than establishing how authority had been negotiated among all involved actors.}

P10 described this shift clearly: \quoteit{I can still see these data, but how they should be handled is decided by my family and doctors---for example, whether to adjust medication dosage.} \revisionzl{She described the system as continuing to address her as if she remained the decision-maker:} \quoteit{I do not make those decisions myself, but my phone keeps telling me about these things.} \revisionzl{We describe this patient-experienced condition as being \textit{visible but not decidable}: the data remained addressed to the participant even though she reported that the relevant decisions and actions had shifted to family members or clinicians.} P4 added a further observation: \quoteit{Nowadays, all decisions regarding my medication are made by my daughter; she monitors my data every day and then tells me what to do. Yet my phone still reminds me daily---almost as if it doesn't realize that I am no longer the one in charge of myself.} \revisionzl{P4 and P10 described redistributed decision responsibility from their perspectives; without caregiver interviews, these findings do not establish how the family members involved understood or negotiated those arrangements.}

\subsection{The Failure of Disengagement}
\label{results:partial-disengagement}

\revisionzl{\revisionzl{Participants described reducing their engagement when data were difficult to understand or no longer supported practical action---for example}, \revisionzl{by closing applications, ignoring notifications, or avoiding bedside monitors}. \revisionzl{Many recalled that these efforts were difficult to sustain because data remained continuously present across ubiquitous systems}. \revisionzl{Rather than fully withdrawing}, participants described partial disengagement: \revisionzl{they temporarily reduced contact with data but often returned to it later}. \revisionzl{We therefore interpret the disengagement participants described as} unstable and limited.} In the following section, we unpack this failure of disengagement in two parts. First, \revisionzl{participants described some kinds of health data as harder to avoid than others}. Second, \revisionzl{they described strategies for attending to some data while ignoring others, but recalled that these strategies were difficult to sustain and that they often returned to data they had tried not to view}.

\noindent \textbf{The Uneven Avoidability of Data.} \revisionzl{Participants described the data types identified in Section}~\ref{results:data-environment} \revisionzl{as varying in how easy they were to avoid, which we interpret as shaping the forms of disengagement available to them}. \revisionzl{They described data that required more interpretation, such as state signals and some trajectories, as continuously present and difficult to escape even briefly}. By contrast, \revisionzl{they described some system-generated interpretations as more avoidable when access required an additional step}.

\revisionzl{Participants described state signals as the hardest data to avoid and reported that they were continuously produced and displayed through bedside monitors, wearables, and phone interfaces, remaining in view even without active checking}. P19 described the bedside monitor as something that was difficult to get away from even without actively checking it: \quoteit{That monitor is always there, and the numbers keep jumping. Even if you do not want to look, it is still there. Sometimes you catch it in the corner of your eye. It is very hard to not notice it completely.} \revisionzl{As participants described it, avoidance did not mean deciding not to open or check something, but trying not to attend to data already present}.

P7 described a similar experience with a wearable: \quoteit{The value on the wristband is always updating, and as soon as the phone lights up you can see it. Sometimes I do not want to look, but it is right there, and it is very easy to tap it open.} \revisionzl{P7 indicated that the device's always-on design made disengagement difficult for her; because she described the data as already in view, avoiding them meant trying to check them less often and devote less attention to them}. Her closing request, \quoteit{Can you help me turn it off?}, \revisionzl{further suggests that she perceived no easy way to step away without help}. \revisionzl{Participants described trajectories and algorithmic outputs as easier to avoid when accessing them required an additional step, such as opening an app, swiping through a notification, or navigating to a particular screen}. P6 described choosing not to look at trajectories, at least for a time: \quoteit{Sometimes I do not really want to look at the curve, because when I look at it I start thinking too much.} P20 described a similar pattern with prompts based on algorithmic outputs: \quoteit{Sometimes I swipe away those prompts without opening them.} \revisionzl{These examples suggest that participants experienced such data as more avoidable when access depended more on their own initiative}.


\noindent\textbf{Selective Attention and Its Instability.} \revisionzl{Many participants described adopting selective engagement when faced with data they felt unable to avoid and did not always want to confront}. \revisionzl{Rather than disengaging completely, they recalled focusing on data that still seemed worth tracking and ignoring the rest}. P21 described this clearly: \quoteit{Now I mainly look at my medication situation, because that is still something I can do. As for the other data, like those prompts or curves, I just do not really bother with them anymore.} \revisionzl{Participants framed this strategy as preserving attention for data that still felt actionable while stepping back from data that no longer did}.

\revisionzl{Both P6 and P20 recalled that this avoidance did not last}. P6 explained, \quoteit{But after a while I still click in to take a look, mainly because I want to know roughly what direction things are going. Otherwise, I do not feel safe.} Similarly, P20 described returning to prompts she had initially dismissed, because she kept wondering \quoteit{whether they were reminding me of something important, and later I go back to check them.} \revision{\revisionzl{They associated returning to these data with seeking reassurance or avoiding missed information}, \revisionzl{illustrating why complete disengagement was} \revisionzl{difficult to sustain.}}

\revisionzl{\revisionzl{Home-based participants described greater ease in disengaging from some data and attributed this partly to the absence of bedside monitors from their reported daily environments}. Nevertheless, these participants \revisionzl{also described} difficulty fully disengaging. For example, P3 explained, \quoteit{Yes, sometimes I wish I could get rid of this bother [data exposure], but it's hard to actually do. It's true I don't have that machine at the bedside constantly telling me things, but actually, on one hand, sometimes I still end up looking at the CGM data and records, and I worry about that too. On the other hand, I'm also curious myself deep down, so avoiding it is always... overall I think the effect is quite limited.}}

\revisionzl{Taken across the two breakdowns and participants' attempts to disengage, iteration was sometimes described as allowing adjustment: P19 described gradually learning that some alerts were less consequential, and P24 reported that restating his condition helped the system adapt somewhat.} \revisionzl{However, we interpret repair as neither automatic nor cost-free.} \revisionzl{P24 described having to re-establish the same context in each interaction, while P6 and P20 described returning to data they had tried to avoid, which they associated with persistent uncertainty.} \revisionzl{These retrospective descriptions do not establish how much time individual encounters displaced, but they illustrate why later repair could not be assumed: subsequent encounters could reproduce interpretive and practical work within a care horizon participants already understood as limited.}

\subsection{\revisionzl{Contextual Conditions Shaping the Breakdowns}}
\label{results:contextual-conditions}
\revisionzl{The analytical lens identifies where breakdowns occurred in participants' engagement with data; our analysis links these breakdowns to three conditions: the technological and care environment, the relationships negotiating interpretation and action, and available institutional resources.}
\noindent\textbf{\revisionzl{Technological and care-environment conditions.}} \revisionzl{Hospital-based participants described bedside monitors, mainly clinician-facing, as continuously visible and hard to refuse; mobile applications and wearables, more often patient-facing, offered more control over whether data were opened, checked, or ignored, though notifications still drew several participants back in. Our hospital-heavy sample does not support a balanced comparison, but disengagement was hardest when clinician-facing data remained present within participants' surroundings, linking disengagement difficulty to a device's intended audience, mode of exposure, and placement in care rather than to a uniform property of every device.}
\noindent\textbf{\revisionzl{Relational and decision-making conditions.}} \revisionzl{Participants described interpretation and action as occurring through relationships with clinicians and family rather than individually. P1 saw repeated questioning as likely to annoy clinicians, and P19 worried that calling a doctor over a non-urgent alert would upset them; both associated a social cost with seeking clarification, constraining access to needed clinical interpretation. P4 and P10 described medication and care decisions as shifted to family or clinicians, while the systems they encountered still addressed them as the person expected to act. This is a mismatch between distributed relationships and systems that model interpretation and action as an individual process, not a problem of shared or transferred authority itself.}
\noindent\textbf{\revisionzl{Institutional and resource conditions.}} \revisionzl{Access to an explanatory pathway depended on clinical interpretation being available and on whether displayed signals supported interpretation at all. P26 cited limited clinical time as restricting explanations, noting isolated fluctuations sometimes lacked stable clinical meaning; P27 added that clinical oversight could not always backstop unclear outputs. The missing pathway thus sits at the intersection of interface limitations, ambiguous signals, and limited clinical access, not system design or patient literacy alone.}
\noindent\textbf{\revisionzl{Analytic roles of patient and clinician interviews.}} \revisionzl{The experiential themes derive from patient interviews; clinician interviews were not a parallel or validating source but related to them in three ways. First, the groups spoke from different standpoints without factually contradicting each other: patients described confusion and guesswork from a lived position, clinicians described device intent and structural constraints from a provider position, as with P28 on bedside data being clinician-oriented (contextualizing P1's inability to read the ECG) and P26 and P27 on limited consultation time and ambiguous signals (contextualizing P1 and P19's reluctance to question doctors). Second, clinician accounts corroborated patient reports as systemic, as when P30's account of long-horizon advice conflicting with end-of-life priorities corroborated the fatigue P3 and P14 described. Third, clinicians and patients occasionally diverged evaluatively: P26 and P27 saw limited explanation and delayed oversight as reasonable given structural constraints, while P1 and P19 experienced the same conditions as unaddressed uncertainty. Clinician interviews were thus convergent and qualifying rather than independently confirmatory, deepening our interpretation of contextual conditions without establishing patients' subjective experiences themselves; we found little factual disagreement between the two groups but do not report this as an agreement rate, given their different analytic roles.}
\revisionzl{These contextual conditions are not additional stages in the data--interpretation--action loop but conditions shaping how the broader breakdowns were described: repeated and unresolved sensemaking was tied to ambient exposure, socially constrained access to explanation, and limited clinical capacity; care-context misalignment became especially visible when decision authority was distributed across patients, families, and clinicians; and disengagement varied by whether data were embedded in clinician-facing infrastructure or accessed through more controllable personal devices. The Discussion returns to which patterns may extend beyond this study and which manifestations relate to Chinese tertiary care.}

\subsection{\revisionzl{Expectations for Data Exposure}: Less Data, More Support}
\label{results:desired-reorientation}
\revisionzl{Participants did not describe wanting to reject data from ubiquitous computing systems}. Instead, they questioned whether support should continue to take the form of more data and more frequent updates. Across interviews, participants described a shift in \revisionzl{their expectations}: they wanted systems to reduce unnecessary data exposure while preserving, or even expanding, forms of support that remained meaningful in the end-of-life context.

\noindent \textbf{Less data: \revisionzl{exposure should be less visible, less frequent, and more selective}.}
\revisionzl{Participants linked their preference for less visible, less frequent, and more selective data presentation to the burden they described in interpreting continuously updated data}. P20 explained, \quoteit{That heart monitor is always there, jumping all the time. In many cases there is actually no meaningful change. I feel that if nothing is especially abnormal, it does not have to keep displaying there all the time. Or could it at least be changed into another form? Looking at it is really terrifying. Or could it be moved to the doctors' room? I cannot understand it anyway, so why let me see it?} Similarly, P18 emphasized \revisionzl{that prompts} should be necessary and doable: \quoteit{Sometimes it pushes suggestions, but a lot of the time they are all pretty similar anyway. I feel that it does not need to remind me so frequently. What matters more is whether it can offer suggestions that are actually doable in my current condition, and avoid suggesting things that do not need to be done. I think that is very important.} \revisionzl{Across these statements, participants} preferred information to appear only when it marked something serious, urgent, or directly relevant. As P8 put it, \quoteit{If something is especially dangerous---if a certain condition appears and you need to pay attention immediately---I hope it can just say that clearly, instead of making me figure it out from all those data myself.}

\noindent\textbf{More support: reorienting data exposure from bodily decline to lived personhood.} \revisionzl{Participants did not describe wanting to remove data exposure from end-of-life care; instead}, they \revisionzl{expressed a preference for} a different kind of exposure. P6 put it clearly: \quoteit{I think what I need most right now is a system that witnesses and accompanies my life, rather than a system that identifies all my bodily indicators. You know, that kind of system is for people who can still be cured.} Therefore, a more meaningful question in the end-of-life context is how data exposure could still help once a cure is no longer the goal. \revisionzl{The interviews suggest one preference: participants valued data exposure when they imagined it as an interactional response rather than a readout}. \revisionzl{We analyze this preference} along two dimensions---the aim of exposure and its form.

\textit{Deepening the aim: from clinical tracking to recognition of inner life.} \revisionzl{Participants described wanting exposure to register what they were going through rather than diagnose or correct it}. P12 proposed an interaction design to enhance her sense of time: \quoteit{I hope that through slowly changing environmental cues, slight periodic signals, or perceptible temporal markers, time would no longer just flow by without difference, but could become something that is experienced and lived through.} \revisionzl{P12 imagined that such cues could help} time feel lived rather than measured. Others described the aim as a quiet acknowledgment. P19 said, \quoteit{It is not always that I want to know what to do. Sometimes I just hope there is something that can acknowledge that right now you are scared, you are tired, you feel like you cannot hold on. At moments like that, it is not about solving a problem, but about something knowing that this is what you are going through.} \revisionzl{P19 thereby expressed a preference for exposure that acknowledges what patients are experiencing without requiring the system to solve a concrete problem.}

\textit{Broadening the form: from on-screen display to ambient, autonomous response.} Participants also \revisionzl{described wanting} exposure to reach them through the environment rather than through information they had to read. P8 described the minimal version: \quoteit{I just hope there is something there. It does not keep disturbing you, but you know it is accompanying you. For example, if it detects some physiological signs of deep sadness, it could just give a simple response. That would already be enough.} P4 gave a more physical example: \quoteit{Sometimes at night the room feels too dark, but I am too tired to get up and turn on the light. If it could brighten itself a little, it would make people feel much better.} P12 drew \revisionzl{a sharp} contrast with instruction: \quoteit{If it keeps telling you what to do every time, that is actually tiring. But if it just adjusts things for you on its own, and you simply feel more comfortable, then that feels much more natural.} \revisionzl{Taken together, these examples suggest that participants wanted data-enabled systems to respond on their behalf through accompaniment or environmental adjustment, without requiring them to interpret data or take action}.

\newcolumntype{L}[1]{>{\raggedright\arraybackslash}p{#1}}

\begin{footnotesize}
\begin{longtable}{L{0.18\linewidth} L{0.20\linewidth} L{0.34\linewidth} L{0.20\linewidth}}
\caption{\revisionzl{Summary of participant-reported experiences and patient-derived themes from 25 patient interviews, contextualized by 8 clinician interviews.} \revisionzl{The themes are organized around data interpretation, action, and the experiences of partial disengagement described by participants.}}
\label{tab:themes} \\
\toprule
\textbf{Theme} & \textbf{Sub-theme} & \textbf{Description} & \textbf{Illustrative Participants} \\
\midrule
\endfirsthead

\multicolumn{4}{l}{\textit{Table~\ref{tab:themes} continued from previous page}} \\
\toprule
\textbf{Theme} & \textbf{Sub-theme} & \textbf{Description} & \textbf{Illustrative Participants} \\
\midrule
\endhead

\midrule
\multicolumn{4}{r}{\textit{Continued on next page}} \\
\endfoot

\bottomrule
\endlastfoot

\multirow{4}{=}{\textbf{Living with data:} End-of-life care as a pervasive health-data environment (\S4.1)}
& State signals & \revisionzl{High-frequency physiological readings (ECG, glucose) that participants described as continuously visible but not always interpreted}. & P11, P3 \\
\cmidrule(l){2-4}
& Trajectories & Trend curves and fluctuation prompts that \revisionzl{participants described as orienting them} to bodily change over time. & P6, P15 \\
\cmidrule(l){2-4}
& Algorithmic outputs & Inferred risk levels, abnormality labels, and recommendation snippets \revisionzl{that participants described receiving} as conclusions. & P9, P18 \\
\cmidrule(l){2-4}
& Behavioral audits & Records of completed or missed actions (medication, diet) \revisionzl{that participants described using} for backward review. & P5, P13 \\
\midrule

\multirow{4}{=}{\textbf{Repeated and unresolved sensemaking:} \revisionzl{data described as visible but difficult to resolve (\S4.2)}}
& Decoding the body via state signals & \revisionzl{Participants described trying to infer the meaning of single readings without medical or temporal context and forming unstable hypotheses}. & P1, P7; clinician P26 \\
\cmidrule(l){2-4}
& Reasoning across time via trajectories & \revisionzl{Participants described extending inference backward and forward from trends without resources to settle it}. & P6, P15, P21 \\
\cmidrule(l){2-4}
& Reconstructing system logic from algorithmic outputs & \revisionzl{Participants described receiving conclusions without supporting evidence and trying to infer which alerts mattered}. & P16, P19 \\
\cmidrule(l){2-4}
& Behavioral audits as the exception & \revisionzl{P5 and P13 described the action--record pathway as comparatively transparent and requiring little further interpretation}. & P5, P13 \\
\midrule

\multirow{3}{=}{\textbf{Misalignment in context management:} \revisionzl{participants described difficulty moving from interpretation to action (\S4.3)}}
& When the body can no longer act & \revisionzl{Participants described suggestions that assumed capacities they no longer had and reminders that marked lost ability}. & P11, P24 \\
\cmidrule(l){2-4}
& When goals no longer apply & \revisionzl{Participants described long-horizon optimization prompts as persisting after they had reprioritized comfort and immediacy}. & P3, P14; clinician P30 \\
\cmidrule(l){2-4}
& \revisionzl{When participants described control as having moved elsewhere} & \revisionzl{Patients reported continuing to receive data after relevant decisions had shifted to family members or clinicians (``visible but not decidable'').} & P4, P10 \\
\midrule

\multirow{2}{=}{\textbf{Failure of disengagement:} unstable, partial withdrawal (\S4.4)}
& Uneven avoidability across data types & \revisionzl{Participants described ambient state signals as harder to escape and some outputs as more avoidable when accessing them required an additional step}. & P7, P19, P20 \\
\cmidrule(l){2-4}
& Selective attention and its instability & \revisionzl{Participants described focusing on still-actionable data, ignoring other data, and later returning amid uncertainty}. & P6, P20, P21 \\
\midrule

\multirow{2}{=}{\textbf{Reorientation:} minimize data, maximize support (\S4.5)}
& Less visible, less frequent, more critical exposure & \revisionzl{Participants preferred serious, urgent, or directly relevant information to be surfaced selectively}. & P8, P18, P20 \\
\cmidrule(l){2-4}
& From bodily readout to existential support & \revisionzl{Participants expressed a preference for ambient, autonomous responses that acknowledged personhood rather than imposing further instruction}. & P4, P6, P8, P12, P19 \\

\end{longtable}
\end{footnotesize}




%% file: sections/discussion.tex
In this section, \revisionzl{we situate ubiquitous exposure to health data within end-of-life care.} \revisionzl{We then discuss how ubiquitous health systems might better support sensemaking,} \revisionzl{enable meaningful data-informed action,} \revisionzl{and address existential needs through the ways they present data.} \revisionzl{As in the Findings, unqualified references to ``participants'' denote patient participants; clinician evidence is identified explicitly when it provides context.}

\subsection{Rethinking Data Exposure in End-of-Life Ubiquitous Computing}

Ubiquitous computing has often treated data exposure as a prerequisite for better awareness and decision-making. In healthcare, this assumption is particularly powerful: \revisionzl{continuous sensing, bedside monitoring, wearables, and mobile applications are expected to make health information more visible and thereby enable patients, caregivers, and clinicians to reflect and respond more effectively.} In end-of-life care, this promise appears even more compelling because bodily change is frequent, uncertainty is high, and decisions can feel urgent. \revisionzl{Drawing on participant interviews, our analysis complicates the assumption that visibility is inherently supportive. Participants linked different forms of ubiquitous health data to different interpretive and practical demands} (Section~\ref{results:data-environment}); \revisionzl{we identified three recurrent breakdown patterns: difficulty forming stable interpretations, difficulty translating interpreted outputs into meaningful action, and difficulty disengaging when data remained embedded in participants' care environments.}

First, as detailed in Section~\ref{results:unresolved-sensemaking}, \revisionzl{data exposure prompted interpretive work that often remained unresolved: across state signals, trajectories, and algorithmic outputs, participants compared current values with prior experiences, searched for plausible causes, and tried to reconstruct why a system had produced a particular output, but these efforts often failed because the displayed outputs lacked the medical, temporal, or causal context needed. We therefore interpret the problem not as low health literacy or insufficient motivation, but as a gap between continuous data availability and support for meaning-making.}

Second, as detailed in Section~\ref{results:care-context-misalignment}, \revisionzl{participants could sometimes interpret an output but not connect it to meaningful action, as systems kept offering suggestions, reminders, and behavioral prompts as though patients remained capable of ongoing self-management.} \revisionzl{Three forms of mismatch emerged: suggested actions that were physically impossible given participants' capacities; actions technically possible but misaligned with their time horizons or care priorities; and outputs that kept addressing patients after relevant decisions had shifted to family members or clinicians.} \revisionzl{An output's actionability thus depended on its fit with} the patient's bodily capacity, care goals, and decision-making role, \revisionzl{and could remain understandable yet cease to support action once these circumstances changed.}

Third, as detailed in Section~\ref{results:partial-disengagement}, \revisionzl{participants described difficulty stepping away from data that no longer helped, checking less often, looking away, dismissing prompts, or ignoring curves, yet returning when uncertainty persisted. Trajectories and algorithmic judgments were often easier to avoid, but participants linked their return to reassurance, risk awareness, or clinical routines, supporting the claim that data embedded in care infrastructures can be difficult to avoid.}

\revisionzl{Each pattern is familiar on its own: personal-informatics research has examined cycles of collection, reflection, and action, as well as burden and disengagement, while context-aware computing has studied mismatches between system assumptions and changing situations}~\cite{li2010stage,epstein2015lived,epstein2016beyond,ancker2015you,schilit1994context}. \revisionzl{The four data forms serve a comparative purpose rather than a universal taxonomy, letting us examine the different demands associated with state signals, trajectories, algorithmic outputs, and behavioral audits. Our contribution is to show how these familiar problems combined at the end of life and became more consequential within patients' limited remaining time.}

\revisionzl{Time mattered because later encounters sometimes allowed adjustment, but that adjustment required further work: P19 learned that some alerts were less consequential, and P24 reported that restating his condition helped the system adapt somewhat. Participants also described revisiting unresolved data, restating context systems failed to retain, receiving recommendations aimed at futures they no longer prioritized, or returning to data they had tried to avoid, showing why systems cannot assume repair will be available or cost-free.} \revisionzl{By \textit{temporal irreversibility}, we mean that a later explanation or better-adapted recommendation may improve what happens next but cannot recover time and attention already spent. Lost time matters more when future opportunities are few and immediate comfort is itself a care priority. Systems should therefore reduce avoidable misalignment when data are presented and allow patients to decline data when the likely benefit does not justify the immediate burden.} The next section examines how technological, relational, and institutional conditions shaped these patterns in this study.


\subsection{\revisionzl{From Broader Patterns to Situated Manifestations}}
\label{discussion:situated}

\revisionzl{The ``data--interpretation--action'' lens identifies where breakdowns occurred but does not imply every manifestation is equally transferable across devices, cultures, or healthcare systems. Our analysis identifies three recurrent interactional patterns: visible data may fail to support a workable interpretation; interpreted data may fail to connect to an action the patient can, wants to, or is authorized to take; and disengagement may remain unstable when data are embedded in care. We interpret their form and prominence as situated in the technological and care environment, patients' relationships with clinicians and family, and the institutional availability of clinical interpretation} (Section~\ref{results:contextual-conditions}).

\noindent\revisionzl{\textbf{Technological and care-environment conditions.}} \revisionzl{Disengagement was hardest around clinician-facing bedside monitors continuously visible in hospital rooms, whereas patient-facing applications and wearables more often required deliberate access or allowed dismissal, though participants still reported returning amid uncertainty. Avoidability may depend on how exposure is configured: the intended user, whether presentation is ambient or user-initiated, and whether the device is embedded in a space the patient can leave or control. Because 21 of 25 patient participants were hospital-based, this is a situated pattern rather than a balanced comparison of hospital and home care.}

\noindent\revisionzl{\textbf{Relational and cultural conditions.}} \revisionzl{The social cost of seeking clarification described by P1 and P19 is not unique to Chinese or East Asian care: patients across healthcare settings may defer to clinicians or avoid questions for fear of being regarded as difficult or jeopardizing the relationship}~\cite{frosch2012authoritarian,joseph2014knowledge}. \revisionzl{Research on Chinese cancer care nevertheless describes physician-directed communication and power asymmetries}~\cite{tu2019outpatient}, \revisionzl{while work on end-of-life decision-making documents family participation in Chinese settings}~\cite{yang2022confucian,bowman2001chinese}. \revisionzl{We therefore read participants' reluctance to question clinicians and reports of transferred decision authority not as inherent Chinese deference, but as relational patterns whose salience may reflect the cultural and institutional conditions of Chinese tertiary care. The breakdown ``when control has moved elsewhere'' thus lies not in family or clinician involvement itself, but in systems continuing to address patients as the sole interpreter and actor after decision relationships had changed.}

\noindent\revisionzl{\textbf{Institutional and resource conditions.}} \revisionzl{The interviews and P26's account suggest the missing explanatory pathway involved displayed data lacking sufficient context, isolated fluctuations lacking stable clinical significance, and limited clinical time.} \revisionzl{Limited consultation time is documented as a barrier to shared decision-making generally}~\cite{legare2008barriers} \revisionzl{and specifically in Chinese care}~\cite{tu2019outpatient,huang2015shared}. \revisionzl{The breakdown thus sits at the intersection of system presentation, signal ambiguity, and clinical workload, not interface design or staffing alone, which limits what design can accomplish: systems may communicate provenance and uncertainty, distinguish answerable questions from those requiring clinical judgment, and route clarification requests, but cannot make indeterminate signals determinate or replace professional judgment.}

\revisionzl{We therefore frame the three breakdown patterns as potentially transferable, but their frequency, intensity, and concrete manifestations as situated findings requiring comparison across devices, care environments, cultures, and healthcare systems.}

\subsection{Designing for \revisionzl{Interpretive Stability}}

\revisionzl{Prior work shows that patient interpretation depends on representational form, contextual information, and opportunities for collaborative sensemaking}~\cite{hohenstein2018supporting,raj2019clinical,chung2016boundary}. \revisionzl{Our analysis examines this problem in end-of-life data encounters: participants described repeatedly trying to explain fluctuations, trajectories, and labels that remained visible without sufficient context} (Section~\ref{results:unresolved-sensemaking}), \revisionzl{while also asking for less frequent exposure and clearer indications of what required attention} (Section~\ref{results:desired-reorientation}). \revisionzl{We use \textit{interpretive stability} to mean the extent to which a representation supports a bounded, provisional understanding---for example, whether a change is isolated or sustained, what the system can and cannot infer from it, and whether it warrants further attention---without repeatedly reopening the same unresolved question.} \revisionzl{It does not mean eliminating clinical uncertainty or replacing professional judgment} (Section~\ref{results:contextual-conditions}).

\revisionzl{For state signals and trajectories, this requires \textit{layered interpretive support} rather than treating the problem as a deficit in patient literacy.} \revisionzl{A first layer could identify what a value represents, its source, and the temporal or reference context available to the system; a second could explain in plain language whether the system can distinguish an expected variation from a potentially consequential change and explicitly state when it cannot}~\cite{hohenstein2018supporting,raj2019clinical}. \revisionzl{A \textit{bounded escalation pathway} begins where system-provided interpretation reaches its limit: patients could record or route an unresolved question, while locally defined urgency criteria could distinguish immediate escalation from questions suitable for later review.} \revisionzl{This proposal responds to several findings: P1 and P19 reported difficulty obtaining an explanation and judging when to seek one; P26 identified signal ambiguity and limited clinical time; P27 noted that applications did not distinguish urgency; and P8 asked that genuinely dangerous conditions be stated clearly.} \revisionzl{Prior work treats patient--provider interpretation as collaborative or socially scaffolded work}~\cite{chung2016boundary,mamykina2008mahi,chen2026not}; \revisionzl{routing must therefore be co-designed with patients and care teams around available clinical capacity so that it does not simply relocate interpretive burden to clinicians.}

\revisionzl{For algorithmic outputs, P16's reported uncertainty about which indicator produced an ``abnormal'' label and P19's reported uncertainty about which alerts signaled real danger support a narrower form of interactive explanation.} \revisionzl{Rather than presenting a closed label, a system could allow patients to inspect which available signals contributed to it, what the output does and does not establish, and whether it warrants urgent attention.} \revisionzl{This proposal builds on prior work on contextual health-data sensemaking and comprehensible result presentation}~\cite{raj2019clinical,hohenstein2018supporting,meng2026tibetcpr,zhang2025slideaudit}, \revisionzl{but remains a design hypothesis to be evaluated with end-of-life patients.} \revisionzl{An explanation should not imply causality or clinical certainty that the underlying data cannot support.}

\revisionzl{The implication for behavioral audits is narrower.} \revisionzl{P5 and P13 described using records of medication, diet, or activity to look back at whether an already understood action had occurred}; \revisionzl{unlike physiological signals or inferred labels, the action--record relation was direct.} \revisionzl{Personal-informatics and patient--provider collaboration research likewise treats such records as resources for reflection and coordination}~\cite{li2010stage,chung2016boundary}. \revisionzl{Systems should therefore preserve that clarity by limiting audit-style representations to actions that can be recorded directly and by keeping them visually distinct from bodily outcomes}. \revisionzl{Our interviews do not support inferring that a recorded action caused a subsequent physiological change; any such association should remain explicitly uncertain rather than being presented as an audit fact}.


\subsection{Designing for Care-Context Alignment}

As detailed in Section~\ref{results:care-context-misalignment}, \revisionzl{participants described action-oriented outputs that no longer fit what their bodies could do, what mattered to them within their remaining time, or who held responsibility for a decision.} \revisionzl{Context-aware computing has long sought to adapt system behavior to changing users and environments}~\cite{schilit1994context,weiser1999computer}, \revisionzl{and personal-informatics research has examined how data are translated into action over time}~\cite{li2010stage,epstein2015lived}. \revisionzl{Our contribution is not a new model of context adaptation, but evidence that downstream prompts retained assumptions participants described as no longer fitting their situations amid changes in bodily capacity, care priorities, and decision relationships. We use \textit{care-context alignment} to describe how well an action-oriented output fits these three dimensions at the time it is presented.} \revisionzl{FHIR(Fast Healthcare Interoperability Resources)-based pipelines may make device-generated and clinical data available across systems}~\cite{abedian2025streamlining,ayaz2021fast,ma2026can,liu2025supporting}, \revisionzl{but interoperability alone does not determine whether an output remains appropriate, beneficial, or addressed to the right person.}

\noindent\textbf{\revisionzl{Aligning outputs with bodily capacity.}} \revisionzl{P11 reported receiving activity reminders after walking had become impossible; P24 had to restate his inability to exercise before a conversational system adapted its advice. These cases extend prior critiques of tracking burden and changing life circumstances}~\cite{ancker2015you,epstein2016beyond}, \revisionzl{locating burden not only in data collection but in prompts that encoded an outdated body and could render lost capacity as failure. A capability-gating mechanism could screen action-oriented prompts against current, reliable capacity information and clinician-specified constraints, deferring, suppressing, or replacing conflicting prompts with a lower-burden alternative while preserving access to clinically necessary information. Because capacity information may be incomplete or stale, the mechanism should expose its source and recency rather than present automated gating as clinical judgment.}

\noindent\textbf{\revisionzl{Aligning outputs with current care priorities.}} \revisionzl{P10 prioritized effects that mattered immediately, P6 interpreted trajectories in relation to remaining time, and P14 kept receiving blood-pressure reminders that no longer mattered to him though his doctors still valued the record. These cases give interactional form to the broader shift in end-of-life care from long-term optimization toward comfort and quality of remaining life}~\cite{sallnow2022report,temel2010early}. \revisionzl{Digital advance-care-planning systems support articulating values and preferences}~\cite{roberts2024personalized,sudore2017effect}; \revisionzl{our analysis motivates carrying such preferences into the logic governing recommendations and data exposure. A care-priority profile should be explicit, revisable, and linked to particular outputs rather than fixed or inferred silently from behavior; a patient-selected ``comfort-first'' priority, for example, could reduce non-urgent lifestyle-optimization prompts while retaining chosen information. The profile should record who entered or revised a priority and remain reviewable by the patient and, when authorized, caregivers and clinicians.}

\noindent\textbf{\revisionzl{Aligning outputs with decision relationships.}} \revisionzl{P4 and P10 reported medication or care decisions shifted to family members or clinicians while prompts kept addressing them as the expected actor. Prior work shows health-data work is distributed across patients, caregivers, and clinicians and negotiated at their boundaries}~\cite{bhat2023we,miller2016partners,chung2016boundary,kaziunas2017caring}. \revisionzl{This recipient--role mismatch motivates a preliminary \textit{role-aware routing} hypothesis: systems could maintain an explicit, editable specification of who receives which output for a given decision, so the same blood-pressure event might appear as plain-language status to the patient, an action prompt to a caregiver, and a clinically contextualized value to a clinician. Roles should be specified per decision, remain visible and reversible, and never remove patient access merely because another person receives an action-oriented version. Because caregivers were not interviewed, this proposal does not establish their preferred roles or willingness to assume additional work; routing rules require caregiver-inclusive co-design and evaluation.}

\revisionzl{These directions may address mismatches but also create risks: cross-system capability data can expand privacy exposure}~\cite{cohen2020ethical,mittelstadt2017ethics,meng2026neglect}; \revisionzl{a care-priority profile can flatten uncertain or changing preferences; and routing can turn a temporary family or clinician role into persistent authority or transfer coordination work to caregivers}~\cite{bhat2023we,miller2016partners,kaziunas2017caring}. \revisionzl{Systems should therefore minimize shared data, display the provenance and freshness of contextual information, keep priorities and roles reviewable, support authorization and revocation, and preserve patient access. An output is not aligned merely because it reaches someone able to act, but only when the context used and responsibility assigned remain legitimate for everyone involved.}




\subsection{Designing for Ubiquitous Existential Support}

\revisionzl{Thanatosensitive HCI has argued that mortality, identity, temporality, and meaning should be treated as constitutive conditions of interaction rather than edge cases}~\cite{massimi2009dying,massimi2011matters,kaptelinin2016making,kaptelinin2018technology}. \revisionzl{While these traditions provide a general orientation, our analysis of the interviews specifies its implications for continuous health-data exposure: participants described how persistent readouts and prompts continued to demand interpretation and action when they sought less exposure, clearer critical information, acknowledgment, lived time, and comfort} (Sections~\ref{results:partial-disengagement} and~\ref{results:desired-reorientation}). \revisionzl{We use \textit{existential support} to describe support for comfort, dignity, acknowledgment, and personhood when bodily optimization is no longer the sole or primary goal.} \revisionzl{From the interviews, we derive two bounded requirements: exposure should be negotiable and reversible, and support should not always require patients to interpret data or initiate action.} \revisionzl{The specific forms through which systems might meet these requirements remain design hypotheses rather than evaluated outcomes.}

\textbf{\revisionzl{Making exposure negotiable and reversible.}} \revisionzl{Prior personal-informatics work has documented burden, lapse, and disengagement}~\cite{epstein2016beyond,ancker2015you}, \revisionzl{while critical data studies show that datafication distributes emotional and relational work}~\cite{kaziunas2017caring,kaziunas2018lived}. \revisionzl{Our analysis identifies an additional infrastructural problem: P7 described wanting help to turn exposure off, and P20 asked for a monitor to be altered or moved.} \revisionzl{P6 and P20 also described returning to data amid persistent uncertainty; participants portrayed clinician-facing monitoring as embedded in care.} \revisionzl{Systems should therefore provide accessible, reversible controls over what patients see and when, without treating withdrawal as abandonment of care or disabling clinically required monitoring.} \revisionzl{How critical exceptions and re-entry should operate requires co-design with patients, caregivers, and clinicians.}

\noindent\textbf{\revisionzl{Supporting low-demand, patient-authorized responses.}} \revisionzl{Prior HCI work on sociotechnical imaginaries in palliative care identifies safe space, uplifting care, connection, and the materiality of care as design concerns}~\cite{ahmadpour2023hci}. \revisionzl{The examples participants provided make this orientation more specific to continuous data exposure: P8 described a simple response to a perceived state, P19 wanted acknowledgment, and P4 and P12 described automatic lighting or temporal cues that could offer comfort without another interpretive task.} \revisionzl{These examples support prototyping patient-authorized, low-demand environmental responses; they do not establish that affect should be inferred automatically or that any particular sensory mapping will be beneficial.} \revisionzl{Such mappings should remain visible, reversible, and co-designed rather than operate as opaque interventions.} \revisionzl{Our evidence does not justify prescribing specific life-review, gamified, or mixed-reality implementations; these remain possible topics for future co-design rather than implications established by this study.}

\revisionzl{One way to operationalize these two requirements is an editable \textit{data-exposure plan}: a patient could specify which data remain visible, how often and in what form they appear, how exposure can be paused or restored, which low-demand environmental responses are permitted, and which urgent exceptions apply.} \revisionzl{Digital advance-care-planning systems already support patients in articulating values and preferences}~\cite{sudore2017effect,roberts2024personalized}; \revisionzl{the contribution here is to extend preference elicitation to the everyday presentation of health data and the authorization of low-demand responses.} \revisionzl{Such a plan is a design hypothesis, not an evaluated intervention, and would require caregiver- and clinician-inclusive co-design to ensure that patient control does not conflict silently with necessary care.}

\revisionzl{These requirements also define the limits of existential support.} Ubiquitous systems should not replace human companionship, simulate intimacy without boundaries, manufacture false hope, aestheticize dying, \revisionzl{or use unbounded affective inference to manipulate patients}. \revisionzl{No prototype was evaluated in this study.} \revisionzl{Drawing on participants' accounts, we argue that evaluations of the withdrawal controls and low-demand ambient responses proposed above should prioritize reductions in interactional burden and the preservation of patients' sense of control, comfort, relational connectedness, and dignity}~\cite{chochinov2008patient,chua2022systematic}. \revisionzl{This study-derived claim is specific to the hospital-dominated, end-of-life context examined here. Its operationalization and validation require participatory, ethically constrained work with patients, caregivers, clinicians, and ethics stakeholders.}

\subsection{Limitations and Future Work}
While this study offers an in-depth \revisionzl{empirical analysis} of data exposure in end-of-life ubiquitous computing, several limitations should be acknowledged.


\revisionzl{First, this study was situated primarily in Chinese tertiary hospitals and did not include a matched cross-cultural or health-system comparison.} As discussed in Section~\ref{discussion:situated}, \revisionzl{the interviews support the identification of relational and institutional conditions that shaped what participants described, but they cannot establish whether clinician deference or family-held decision authority is more prevalent in China than elsewhere, nor can they causally separate cultural norms from hospital workflow and resource scarcity.} \revisionzl{Comparative research across cultural and healthcare settings is needed to examine which manifestations vary and whether the broader breakdown patterns recur.}

\revisionzl{Second, although clinicians were sampled to contextualize clinical interpretation, workflow, and resource constraints, we did not collect a parallel perspective from family caregivers or guardians. As discussed in §3.1, this reflected the IRB's concern about disclosing private, kinship-related information and the phenomenological scope of our research question, which centers on patients' own subjective experience. This did not capture how authority was negotiated, whether caregivers understood those relationships in the same way, or what informational and coordination burdens routing might place on them. We therefore limit "when control has moved elsewhere" to a patient-reported experience rather than a characterization of family-centered decision-making as a whole. The proposed role-aware routing remains a design hypothesis that requires caregiver-inclusive interviews and participatory evaluation before its feasibility or effects can be claimed.}

Third, participants \revisionzl{reported encountering} a heterogeneous set of data-generating systems, including bedside monitors, wearable devices, mobile health applications, and hospital-facing systems. This heterogeneity allowed us to examine data exposure as a broader everyday environment rather than as an interaction with a specific ubiquitous computing system. At the same time, it limits our ability to evaluate the design features or algorithmic mechanisms of any particular system. \revisionzl{Moreover, because 21 of the 25 patient participants were hospital-based, the differences we describe between hospital and home exposure are descriptive rather than a balanced comparison.} Future work could examine specific systems or interface designs to understand how particular forms of data presentation shape interpretation, action, and disengagement.

Fourth, our exclusion criteria also limit the scope of our findings. We excluded individuals whose strong resistance to digital technologies resulted in no relevant prior experience, as well as those with severe hearing or speech impairments that prevented interview participation. Our findings therefore may not capture pre-use technology refusal or the distinctive accessibility barriers experienced by patients with communication impairments.

Fifth, our study relies on semi-structured interviews. \revisionzl{Interviews captured participants' retrospective recollections of data exposure but could not directly observe the moment-by-moment dynamics of noticing a value, comparing it with prior data, or deciding whether to act.} \revisionzl{We did not collect concurrent diaries, screenshots, in-situ observations, experience-sampling data, or interaction logs; our empirical claims therefore concern recurrent patterns described by participants rather than the exact temporal order, duration, or causal unfolding of individual encounters.} \revisionzl{Future work could combine interviews with these forms of in-situ evidence to examine how the ``data--interpretation--action'' loop unfolds over time.}

Finally, \revisionzl{the design implications proposed here require careful validation in end-of-life care.} \revisionzl{They also entail potential risks:} explaining data may create new anxiety; filtering data may hide information that patients or clinicians value; and routing interpretation requests may redistribute responsibility among patients, families, and care teams. Future work should therefore use participatory and ethically constrained design methods to evaluate these ideas with patients, caregivers, clinicians, and ethics stakeholders.

\section{Conclusion}
\revisionzl{This paper examined how terminal-stage patients described living with health data in end-of-life care. We identified three patterns: repeated failed attempts to interpret data lacking context; suggested actions no longer fitting bodily capacities, care priorities, or decision roles; and difficulty sustaining disengagement when data remained embedded in care. These familiar problems take a distinct form at the end of life, where later repair cannot be assumed or cost-free. Patients wanted less visible, less frequent data highlighting critical information, and support centered on comfort and personhood rather than bodily optimization. We argue for systems supporting interpretive stability, care-context alignment, and existential support, evaluated by whether patients understand what matters, act feasibly, decline unwanted exposure, and feel supported without feeling monitored.}

%% file: data_exp.bib
@String{Computing = "Computing" }

@String{Computer = "{IEEE} Computer" }

@String{Springer = "Springer-Verlag" }

@article{drew2014insights,
  title={Insights into the problem of alarm fatigue with physiologic monitor devices: a comprehensive observational study of consecutive intensive care unit patients},
  author={Drew, Barbara J and Harris, Patricia and Z{\`e}gre-Hemsey, Jessica K and Mammone, Tina and Schindler, Daniel and Salas-Boni, Rebeca and Bai, Yong and Tinoco, Adelita and Ding, Quan and Hu, Xiao},
  journal={PloS one},
  volume={9},
  number={10},
  pages={e110274},
  year={2014},
  publisher={Public Library of Science San Francisco, USA}
}

@inproceedings{li2010stage,
  title={A stage-based model of personal informatics systems},
  author={Li, Ian and Dey, Anind and Forlizzi, Jodi},
  booktitle={Proceedings of the SIGCHI conference on human factors in computing systems},
  pages={557--566},
  year={2010}
}

@article{braun2021saturate,
  title={To saturate or not to saturate? Questioning data saturation as a useful concept for thematic analysis and sample-size rationales},
  author={Braun, Virginia and Clarke, Victoria},
  journal={Qualitative research in sport, exercise and health},
  volume={13},
  number={2},
  pages={201--216},
  year={2021},
  publisher={Taylor \& Francis}
}

@article{joseph2014knowledge,
  title={Knowledge is not power for patients: a systematic review and thematic synthesis of patient-reported barriers and facilitators to shared decision making},
  author={Joseph-Williams, Natalie and Elwyn, Glyn and Edwards, Adrian},
  journal={Patient education and counseling},
  volume={94},
  number={3},
  pages={291--309},
  year={2014},
  publisher={Elsevier}
}

@article{tu2019outpatient,
  title={Outpatient communication patterns in a cancer hospital in China: A qualitative study of doctor--patient encounters},
  author={Tu, Jiong and Kang, Ge and Zhong, Jiudi and Cheng, Yu},
  journal={Health Expectations},
  volume={22},
  number={3},
  pages={594--603},
  year={2019},
  publisher={Wiley Online Library}
}

@article{legare2008barriers,
  title={Barriers and facilitators to implementing shared decision-making in clinical practice: update of a systematic review of health professionals’ perceptions},
  author={L{\'e}gar{\'e}, France and Ratt{\'e}, St{\'e}phane and Gravel, Karine and Graham, Ian D},
  journal={Patient education and counseling},
  volume={73},
  number={3},
  pages={526--535},
  year={2008},
  publisher={Elsevier}
}

@article{huang2015shared,
  title={Shared decision-making in the People’s Republic of China: current status and future directions},
  author={Huang, Rongchong and Gionfriddo, Michael R and Zhang, Lizhi and Leppin, Aaron L and Ting, Henry H and Montori, Victor M},
  journal={Patient preference and adherence},
  pages={1129--1141},
  year={2015},
  publisher={Taylor \& Francis}
}

@article{bowman2001chinese,
  title={Chinese seniors’ perspectives on end-of-life decisions},
  author={Bowman, Kerry W and Singer, Peter A},
  journal={Social science \& medicine},
  volume={53},
  number={4},
  pages={455--464},
  year={2001},
  publisher={Elsevier}
}

@article{frosch2012authoritarian,
  title={Authoritarian physicians and patients’ fear of being labeled ‘difficult’among key obstacles to shared decision making},
  author={Frosch, Dominick L and May, Suepattra G and Rendle, Katharine AS and Tietbohl, Caroline and Elwyn, Glyn},
  journal={Health affairs},
  volume={31},
  number={5},
  pages={1030--1038},
  year={2012}
}

@article{braun2021can,
  title={Can I use TA? Should I use TA? Should I not use TA? Comparing reflexive thematic analysis and other pattern-based qualitative analytic approaches},
  author={Braun, Virginia and Clarke, Victoria},
  journal={Counselling and psychotherapy research},
  volume={21},
  number={1},
  pages={37--47},
  year={2021},
  publisher={Wiley Online Library}
}

@article{braun2021one,
  title={One size fits all? What counts as quality practice in (reflexive) thematic analysis?},
  author={Braun, Virginia and Clarke, Victoria},
  journal={Qualitative research in psychology},
  volume={18},
  number={3},
  pages={328--352},
  year={2021},
  publisher={Taylor \& Francis}
}

@article{braun2019reflecting,
  title={Reflecting on reflexive thematic analysis},
  author={Braun, Virginia and Clarke, Victoria},
  journal={Qualitative research in sport, exercise and health},
  volume={11},
  number={4},
  pages={589--597},
  year={2019},
  publisher={Taylor \& Francis}
}

@article{bhat2023we,
  title={" We are half-doctors": family caregivers as boundary actors in chronic disease management},
  author={Bhat, Karthik S and Hall, Amanda K and Kuo, Tiffany and Kumar, Neha},
  journal={Proceedings of the ACM on Human-Computer Interaction},
  volume={7},
  number={CSCW1},
  pages={1--29},
  year={2023},
  publisher={ACM New York, NY, USA}
}

@inproceedings{miller2016partners,
  title={Partners in care: design considerations for caregivers and patients during a hospital stay},
  author={Miller, Andrew D and Mishra, Sonali R and Kendall, Logan and Haldar, Shefali and Pollack, Ari H and Pratt, Wanda},
  booktitle={Proceedings of the 19th ACM Conference on Computer-Supported Cooperative Work \& Social Computing},
  pages={756--769},
  year={2016}
}

@article{senathirajah2014clinician,
  title={The clinician in the Driver’s Seat: Part 1--A drag/drop user-composable electronic health record platform},
  author={Senathirajah, Yalini and Bakken, Suzanne and Kaufman, David},
  journal={Journal of Biomedical Informatics},
  volume={52},
  pages={165--176},
  year={2014},
  publisher={Elsevier}
}

@article{ancker2015you,
  title={“You get reminded you’re a sick person”: personal data tracking and patients with multiple chronic conditions},
  author={Ancker, Jessica S and Witteman, Holly O and Hafeez, Baria and Provencher, Thierry and Van de Graaf, Mary and Wei, Esther},
  journal={Journal of medical Internet research},
  volume={17},
  number={8},
  pages={e202},
  year={2015},
  publisher={JMIR Publications Inc. Toronto, Canada}
}

@article{kaziunas2018lived,
  title={Lived data: tinkering with bodies, code, and care work},
  author={Kaziunas, Elizabeth and Lindtner, Silvia and Ackerman, Mark S and Lee, Joyce M},
  journal={Human--Computer Interaction},
  volume={33},
  number={1},
  pages={49--92},
  year={2018},
  publisher={Taylor \& Francis}
}

@article{basch2017overall,
  title={Overall survival results of a trial assessing patient-reported outcomes for symptom monitoring during routine cancer treatment},
  author={Basch, Ethan and Deal, Allison M and Dueck, Amylou C and Scher, Howard I and Kris, Mark G and Hudis, Clifford and Schrag, Deborah},
  journal={Jama},
  volume={318},
  number={2},
  pages={197--198},
  year={2017}
}

@inproceedings{bardram2006awaremedia,
  title={AwareMedia: a shared interactive display supporting social, temporal, and spatial awareness in surgery},
  author={Bardram, Jakob E and Hansen, Thomas R and Soegaard, Mads},
  booktitle={Proceedings of the 2006 20th anniversary conference on Computer supported cooperative work},
  pages={109--118},
  year={2006}
}

@article{temel2010early,
  title={Early palliative care for patients with metastatic non--small-cell lung cancer},
  author={Temel, Jennifer S and Greer, Joseph A and Muzikansky, Alona and Gallagher, Emily R and Admane, Sonal and Jackson, Vicki A and Dahlin, Constance M and Blinderman, Craig D and Jacobsen, Juliet and Pirl, William F and others},
  journal={New England journal of medicine},
  volume={363},
  number={8},
  pages={733--742},
  year={2010},
  publisher={Mass Medical Soc}
}

@article{sallnow2022report,
  title={Report of the Lancet Commission on the Value of Death: bringing death back into life},
  author={Sallnow, Libby and Smith, Richard and Ahmedzai, Sam H and Bhadelia, Afsan and Chamberlain, Charlotte and Cong, Yali and Doble, Brett and Dullie, Luckson and Durie, Robin and Finkelstein, Eric A and others},
  journal={The Lancet},
  volume={399},
  number={10327},
  pages={837--884},
  year={2022},
  publisher={Elsevier}
}

@article{mamykina2015adopting,
  title={Adopting the sensemaking perspective for chronic disease self-management},
  author={Mamykina, Lena and Smaldone, Arlene M and Bakken, Suzanne R},
  journal={Journal of biomedical informatics},
  volume={56},
  pages={406--417},
  year={2015},
  publisher={Elsevier}
}

@inproceedings{epstein2015lived,
  title={A lived informatics model of personal informatics},
  author={Epstein, Daniel A and Ping, An and Fogarty, James and Munson, Sean A},
  booktitle={Proceedings of the 2015 ACM international joint conference on pervasive and ubiquitous computing},
  pages={731--742},
  year={2015}
}

@inproceedings{homewood2023self,
  title={Self-tracking to do less: An autoethnography of long COVID that informs the design of pacing technologies},
  author={Homewood, Sarah},
  booktitle={Proceedings of the 2023 CHI Conference on Human Factors in Computing Systems},
  pages={1--14},
  year={2023}
}

@inproceedings{satchell2009beyond,
  title={Beyond the user: use and non-use in HCI},
  author={Satchell, Christine and Dourish, Paul},
  booktitle={Proceedings of the 21st annual conference of the Australian computer-human interaction special interest group: Design: Open 24/7},
  pages={9--16},
  year={2009}
}

@inproceedings{epstein2016beyond,
  title={Beyond abandonment to next steps: understanding and designing for life after personal informatics tool use},
  author={Epstein, Daniel A and Caraway, Monica and Johnston, Chuck and Ping, An and Fogarty, James and Munson, Sean A},
  booktitle={Proceedings of the 2016 CHI conference on human factors in computing systems},
  pages={1109--1113},
  year={2016}
}

@inproceedings{rooksby2014personal,
  title={Personal tracking as lived informatics},
  author={Rooksby, John and Rost, Mattias and Morrison, Alistair and Chalmers, Matthew},
  booktitle={Proceedings of the SIGCHI conference on human factors in computing systems},
  pages={1163--1172},
  year={2014}
}

@article{yasmin2020influence,
  title={The influence of mobile phone-based health reminders on patient adherence to medications and healthy lifestyle recommendations for effective management of diabetes type 2: a randomized control trial in Dhaka, Bangladesh},
  author={Yasmin, Farzana and Nahar, Nazmun and Banu, Bilkis and Ali, Liaquat and Sauerborn, Rainer and Souares, Aur{\'e}lia},
  journal={BMC Health Services Research},
  volume={20},
  number={1},
  pages={520},
  year={2020},
  publisher={Springer}
}

@article{epstein2020mapping,
  title={Mapping and taking stock of the personal informatics literature},
  author={Epstein, Daniel A and Caldeira, Clara and Figueiredo, Mayara Costa and Lu, Xi and Silva, Lucas M and Williams, Lucretia and Lee, Jong Ho and Li, Qingyang and Ahuja, Simran and Chen, Qiuer and others},
  journal={Proceedings of the ACM on Interactive, Mobile, Wearable and Ubiquitous Technologies},
  volume={4},
  number={4},
  pages={1--38},
  year={2020},
  publisher={ACM New York, NY, USA}
}

@inproceedings{schilit1994context,
  title={Context-aware computing applications},
  author={Schilit, Bill and Adams, Norman and Want, Roy},
  booktitle={1994 first workshop on mobile computing systems and applications},
  pages={85--90},
  year={1994},
  organization={IEEE}
}

@article{weiser1999computer,
  title={The computer for the 21st century},
  author={Weiser, Mark},
  journal={ACM SIGMOBILE mobile computing and communications review},
  volume={3},
  number={3},
  pages={3--11},
  year={1999},
  publisher={ACM New York, NY, USA}
}

@article{raj2019clinical,
  title={Clinical data in context: towards sensemaking tools for interpreting personal health data},
  author={Raj, Shriti and Lee, Joyce M and Garrity, Ashley and Newman, Mark W},
  journal={Proceedings of the ACM on Interactive, Mobile, Wearable and Ubiquitous Technologies},
  volume={3},
  number={1},
  pages={1--20},
  year={2019},
  publisher={ACM New York, NY, USA}
}

@article{yang2022review,
  title={A review of IoT-enabled mobile healthcare: technologies, challenges, and future trends},
  author={Yang, Yilin and Wang, Haocong and Jiang, Ruizhe and Guo, Xiaonan and Cheng, Jerry and Chen, Yingying},
  journal={IEEE Internet of Things Journal},
  volume={9},
  number={12},
  pages={9478--9502},
  year={2022},
  publisher={IEEE}
}

@article{saleemi2023ubiquitous,
  title={Ubiquitous healthcare: a systematic mapping study},
  author={Saleemi, Maria and Anjum, Maria and Rehman, Mariam},
  journal={Journal of Ambient Intelligence and Humanized Computing},
  volume={14},
  number={5},
  pages={5021--5046},
  year={2023},
  publisher={Springer}
}

@inproceedings{mamykina2008mahi,
  title={MAHI: investigation of social scaffolding for reflective thinking in diabetes management},
  author={Mamykina, Lena and Mynatt, Elizabeth and Davidson, Patricia and Greenblatt, Daniel},
  booktitle={Proceedings of the SIGCHI conference on human factors in computing systems},
  pages={477--486},
  year={2008}
}

@article{wong2015send,
  title={SEND: a system for electronic notification and documentation of vital sign observations},
  author={Wong, David and Bonnici, Timothy and Knight, Julia and Morgan, Lauren and Coombes, Paul and Watkinson, Peter},
  journal={BMC medical informatics and decision making},
  volume={15},
  number={1},
  pages={68},
  year={2015},
  publisher={Springer}
}

@article{tentori2012pervasive,
  title={Pervasive computing for hospital, chronic, and preventive care},
  author={Tentori, Monica and Hayes, Gillian R and Reddy, Madhu},
  journal={Foundations and Trends{\textregistered} in Human--Computer Interaction},
  volume={5},
  number={1},
  pages={1--95},
  year={2012},
  publisher={Emerald Publishing Limited}
}

@article{bardram2005mobility,
  title={Mobility work: The spatial dimension of collaboration at a hospital},
  author={Bardram, Jakob E and Bossen, Claus},
  journal={Computer supported cooperative work (CSCW)},
  volume={14},
  number={2},
  pages={131--160},
  year={2005},
  publisher={Springer}
}

@ArtifactSoftware{R,
    title = {R: A Language and Environment for Statistical Computing},
    author = {{R Core Team}},
    organization = {R Foundation for Statistical Computing},
    address = {Vienna, Austria},
    year = {2019},
    url = {https://www.R-project.org/},
}

@article{lamas2018advance,
  title={Advance care planning documentation in electronic health records: current challenges and recommendations for change},
  author={Lamas, Daniela and Panariello, Natalie and Henrich, Natalie and Hammes, Bernard and Hanson, Laura C and Meier, Diane E and Guinn, Nancy and Corrigan, Janet and Hubber, Sean and Luetke-Stahlman, Hannah and others},
  journal={Journal of palliative medicine},
  volume={21},
  number={4},
  pages={522--528},
  year={2018},
  publisher={SAGE Publications Sage CA: Los Angeles, CA}
}

@article{dillon2017provider,
  title={Provider perspectives on advance care planning documentation in the electronic health record: the experience of primary care providers and specialists using advance health-care directives and physician orders for life-sustaining treatment},
  author={Dillon, Ellis and Chuang, Judith and Gupta, Atul and Tapper, Sharon and Lai, Steve and Yu, Peter and Ritchie, Christine and Tai-Seale, Ming},
  journal={American Journal of Hospice and Palliative Medicine{\textregistered}},
  volume={34},
  number={10},
  pages={918--924},
  year={2017},
  publisher={SAGE Publications Sage CA: Los Angeles, CA}
}

@article{lorenz2008evidence,
  title={Evidence for improving palliative care at the end of life: a systematic review},
  author={Lorenz, Karl A and Lynn, Joanne and Dy, Sydney M and Shugarman, Lisa R and Wilkinson, Anne and Mularski, Richard A and Morton, Sally C and Hughes, Ronda G and Hilton, Lara K and Maglione, Margaret and others},
  journal={Annals of internal medicine},
  volume={148},
  number={2},
  pages={147--159},
  year={2008},
  publisher={American College of Physicians}
}

@article{dai2026exploring,
  title={Exploring Self-Tracking Practices of Older Adults with CVD to Inform the Design of LLM-Enabled Health Data Sensemaking},
  author={Dai, Duosi and Pakianathan, Pavithren VS and Treff, Gunnar and Sareban, Mahdi and Smeddinck, Jan David and Kuoppam{\"a}ki, Sanna},
  journal={arXiv preprint arXiv:2603.23733},
  year={2026}
}

@article{hohenstein2018supporting,
  title={Supporting accurate interpretation of self-administered medical test results for mobile health: assessment of design, demographics, and health condition},
  author={Hohenstein, Jess C and Baumer, Eric PS and Reynolds, Lindsay and Murnane, Elizabeth L and O'Dell, Dakota and Lee, Seoho and Guha, Shion and Qi, Yu and Rieger, Erin and Gay, Geri},
  journal={JMIR Human Factors},
  volume={5},
  number={1},
  pages={e9},
  year={2018},
  publisher={JMIR Publications Toronto, Canada}
}

@article{andersen2011designing,
  title={Designing for collaborative interpretation in telemonitoring: Re-introducing patients as diagnostic agents},
  author={Andersen, Tariq and Bj{\o}rn, Pernille and Kensing, Finn and Moll, Jonas},
  journal={International journal of medical informatics},
  volume={80},
  number={8},
  pages={e112--e126},
  year={2011},
  publisher={Elsevier}
}

@article{suija2022experiences,
  title={Experiences and expectation with the use of health data: a qualitative interview study in primary care},
  author={Suija, Kadri and Mardo, Laura Alexandra and Laidoja, Reet and Nahkur, Siim and Parvelo, Anu and Kalda, Ruth},
  journal={BMC Primary Care},
  volume={23},
  number={1},
  pages={159},
  year={2022},
  publisher={Springer}
}

@article{mcalearney2019patients,
  title={Patients’ perceptions of portal use across care settings: qualitative study},
  author={McAlearney, Ann Scheck and Sieck, Cynthia J and Gaughan, Alice and Fareed, Naleef and Volney, Jaclyn and Huerta, Timothy R},
  journal={Journal of medical Internet research},
  volume={21},
  number={6},
  pages={e13126},
  year={2019},
  publisher={JMIR Publications Toronto, Canada}
}

@article{tieu2015barriers,
  title={Barriers and facilitators to online portal use among patients and caregivers in a safety net health care system: a qualitative study},
  author={Tieu, Lina and Sarkar, Urmimala and Schillinger, Dean and Ralston, James D and Ratanawongsa, Neda and Pasick, Rena and Lyles, Courtney R},
  journal={Journal of medical Internet research},
  volume={17},
  number={12},
  pages={e275},
  year={2015},
  publisher={JMIR Publications Inc. Toronto, Canada}
}

@article{teno2004family,
  title={Family perspectives on end-of-life care at the last place of care},
  author={Teno, Joan M and Clarridge, Brian R and Casey, Virginia and Welch, Lisa C and Wetle, Terrie and Shield, Renee and Mor, Vincent},
  journal={Jama},
  volume={291},
  number={1},
  pages={88--93},
  year={2004},
  publisher={American Medical Association}
}

@article{singer1999quality,
  title={Quality end-of-life care: patients' perspectives},
  author={Singer, Peter A and Martin, Douglas K and Kelner, Merrijoy},
  journal={Jama},
  volume={281},
  number={2},
  pages={163--168},
  year={1999},
  publisher={American Medical Association}
}

@article{sudore2017effect,
  title={Effect of the PREPARE website vs an easy-to-read advance directive on advance care planning documentation and engagement among veterans: a randomized clinical trial},
  author={Sudore, Rebecca L and Boscardin, John and Feuz, Mariko A and McMahan, Ryan D and Katen, Mary T and Barnes, Deborah E},
  journal={JAMA internal medicine},
  volume={177},
  number={8},
  pages={1102--1109},
  year={2017}
}

@article{roberts2024personalized,
  title={A personalized and interactive web-based advance care planning intervention for older adults (Koda Health): pilot feasibility study},
  author={Roberts, R Lynae and Cherry, Katelin D and Mohan, Desh P and Statler, Tiffany and Kirkendall, Eric and Moses, Adam and McCraw, Jennifer and Brown III, Andrew E and Fofanova, Tatiana Y and Gabbard, Jennifer},
  journal={JMIR aging},
  volume={7},
  pages={e54128},
  year={2024}
}

@article{ayaz2021fast,
  title={The Fast Health Interoperability Resources (FHIR) standard: systematic literature review of implementations, applications, challenges and opportunities},
  author={Ayaz, Muhammad and Pasha, Muhammad F and Alzahrani, Mohammed Y and Budiarto, Rahmat and Stiawan, Deris},
  journal={JMIR medical informatics},
  volume={9},
  number={7},
  pages={e21929},
  year={2021},
  publisher={JMIR Publications Toronto, Canada}
}

@article{abedian2025streamlining,
  title={Streamlining wearable data integration for EHDS: a case study on advancing healthcare interoperability using Garmin devices and FHIR},
  author={Abedian, Somayeh and Yesakov, Eugene and Ostrovskiy, Stanislav and Hussein, Rada},
  journal={Frontiers in Digital Health},
  volume={7},
  pages={1636775},
  year={2025},
  publisher={Frontiers}
}

@article{mcgillion2018postoperative,
  title={Postoperative remote automated monitoring: need for and state of the science},
  author={McGillion, Michael H and Duceppe, Emmanuelle and Allan, Katherine and Marcucci, Maura and Yang, Stephen and Johnson, Ana P and Ross-Howe, Sara and Peter, Elizabeth and Scott, Ted and Ouellette, Carley and others},
  journal={Canadian Journal of Cardiology},
  volume={34},
  number={7},
  pages={850--862},
  year={2018},
  publisher={Elsevier}
}

@article{tarbi2025telehealth,
  title={Telehealth for the study of palliative care communication: opportunities, methodological challenges, and recommendations},
  author={Tarbi, Elise C and Schuler, Susanna L and Ambrose, Natalie and Hutchinson, Rebecca N and Reblin, Maija and Cheung, Katharine L},
  journal={BMC palliative care},
  volume={24},
  number={1},
  pages={55},
  year={2025},
  publisher={Springer}
}

@article{krieckemans2025telehealth,
  title={Telehealth in palliative care settings: A systematic review of argument-based ethics literature},
  author={Krieckemans, Leen and Cavolo, Alice and Casaer, Michael and Gastmans, Chris},
  journal={Palliative medicine},
  pages={02692163251360115},
  year={2025},
  publisher={SAGE Publications Sage UK: London, England}
}

@article{lundereng2023health,
  title={Health care professionals’ experiences and perspectives on using telehealth for home-based palliative care: scoping review},
  author={Lundereng, Elias David and Nes, Andrea Aparecida Goncalves and Holmen, Heidi and Winger, Anette and Thygesen, Hilde and J{\o}ranson, Nina and Borge, Christine R{\aa}heim and Dajani, Olav and Mariussen, Kari L and Steindal, Simen A},
  journal={Journal of Medical Internet Research},
  volume={25},
  pages={e43429},
  year={2023},
  publisher={JMIR Publications Toronto, Canada}
}

@article{oechsle2019current,
  title={Current advances in palliative \& hospice care: Problems and needs of relatives and family caregivers during palliative and hospice care—An overview of current literature},
  author={Oechsle, Karin},
  journal={Medical Sciences},
  volume={7},
  number={3},
  pages={43},
  year={2019},
  publisher={MDPI}
}

@article{hebdon2023hospice,
  title={Hospice family caregivers' uncertainty, burden, and unmet needs in prospective audio diaries},
  author={Hebdon, Megan C Thomas and Cloyes, Kristin G and Vega, Marilisa and Rosenkranz, Susan J and Reblin, Maija and Tay, Djin and Mooney, Kathi and Ellington, Lee},
  journal={Journal of Hospice \& Palliative Nursing},
  volume={25},
  number={6},
  pages={321--329},
  year={2023},
  publisher={LWW}
}

@article{beil2019ethical,
  title={Ethical considerations about artificial intelligence for prognostication in intensive care},
  author={Beil, Michael and Proft, Ingo and Van Heerden, Daniel and Sviri, Sigal and Van Heerden, Peter Vernon},
  journal={Intensive care medicine experimental},
  volume={7},
  number={1},
  pages={70},
  year={2019},
  publisher={Springer}
}

@incollection{massimi2009dying,
  title={Dying, death, and mortality: towards thanatosensitivity in HCI},
  author={Massimi, Michael and Charise, Andrea},
  booktitle={CHI'09 Extended Abstracts on Human Factors in Computing Systems},
  pages={2459--2468},
  year={2009}
}

@inproceedings{kaptelinin2016making,
  title={Making the case for an existential perspective in HCI research on mortality and death},
  author={Kaptelinin, Victor},
  booktitle={Proceedings of the 2016 CHI Conference Extended Abstracts on Human Factors in Computing Systems},
  pages={352--364},
  year={2016}
}

@inproceedings{massimi2011matters,
  title={Matters of life and death: locating the end of life in lifespan-oriented HCI research},
  author={Massimi, Michael and Odom, William and Banks, Richard and Kirk, David},
  booktitle={Proceedings of the sigchi conference on human factors in computing systems},
  pages={987--996},
  year={2011}
}

@inproceedings{kaptelinin2018technology,
  title={Technology and the givens of existence: Toward an existential inquiry framework in HCI research},
  author={Kaptelinin, Victor},
  booktitle={Proceedings of the 2018 CHI conference on human factors in computing systems},
  pages={1--14},
  year={2018}
}

@article{chochinov2008patient,
  title={The patient dignity inventory: a novel way of measuring dignity-related distress in palliative care},
  author={Chochinov, Harvey Max and Hassard, Thomas and McClement, Susan and Hack, Thomas and Kristjanson, Linda J and Harlos, Mike and Sinclair, Shane and Murray, Alison},
  journal={Journal of pain and symptom management},
  volume={36},
  number={6},
  pages={559--571},
  year={2008},
  publisher={Elsevier}
}

@article{chua2022systematic,
  title={A systematic scoping review on patients’ perceptions of dignity},
  author={Chua, Keith Zi Yuan and Quah, Elaine Li Ying and Lim, Yun Xue and Goh, Chloe Keyi and Lim, Jieyu and Wan, Darius Wei Jun and Ong, Simone Meiqi and Chong, Chi Sum and Yeo, Kennan Zhi Guang and Goh, Laura Shih Hui and others},
  journal={BMC Palliative Care},
  volume={21},
  number={1},
  pages={118},
  year={2022},
  publisher={Springer}
}

@article{mittelstadt2017ethics,
  title={Ethics of the health-related internet of things: a narrative review},
  author={Mittelstadt, Brent},
  journal={Ethics and Information Technology},
  volume={19},
  number={3},
  pages={157--175},
  year={2017},
  publisher={Springer}
}

@article{basch2016symptom,
  title={Symptom monitoring with patient-reported outcomes during routine cancer treatment: a randomized controlled trial},
  author={Basch, Ethan and Deal, Allison M and Kris, Mark G and Scher, Howard I and Hudis, Clifford A and Sabbatini, Paul and Rogak, Lauren and Bennett, Antonia V and Dueck, Amylou C and Atkinson, Thomas M and others},
  journal={Journal of clinical oncology},
  volume={34},
  number={6},
  pages={557--565},
  year={2016},
  publisher={American Society of Clinical Oncology}
}

@inproceedings{kaziunas2017caring,
  title={Caring through data: Attending to the social and emotional experiences of health datafication},
  author={Kaziunas, Elizabeth and Ackerman, Mark S and Lindtner, Silvia and Lee, Joyce M},
  booktitle={Proceedings of the 2017 ACM Conference on Computer Supported Cooperative Work and Social Computing},
  pages={2260--2272},
  year={2017}
}

@article{cohen2020ethical,
  title={Ethical and legal implications of remote monitoring of medical devices},
  author={Cohen, I Glenn and Gerke, Sara and Kramer, Daniel B},
  journal={The Milbank Quarterly},
  volume={98},
  number={4},
  pages={1257--1289},
  year={2020},
  publisher={Wiley Online Library}
}

@inproceedings{chung2016boundary,
  title={Boundary negotiating artifacts in personal informatics: patient-provider collaboration with patient-generated data},
  author={Chung, Chia-Fang and Dew, Kristin and Cole, Allison and Zia, Jasmine and Fogarty, James and Kientz, Julie A and Munson, Sean A},
  booktitle={Proceedings of the 19th ACM conference on computer-supported cooperative work \& social computing},
  pages={770--786},
  year={2016}
}

@article{veinot2018good,
  title={Good intentions are not enough: how informatics interventions can worsen inequality},
  author={Veinot, Tiffany C and Mitchell, Hannah and Ancker, Jessica S},
  journal={Journal of the American Medical Informatics Association},
  volume={25},
  number={8},
  pages={1080--1088},
  year={2018},
  publisher={Oxford University Press}
}

@article{hoek2017effect,
  title={The effect of weekly specialist palliative care teleconsultations in patients with advanced cancer--a randomized clinical trial},
  author={Hoek, Patrick D and Schers, Henk J and Bronkhorst, Ewald M and Vissers, Kris CP and Hasselaar, Jeroen GJ},
  journal={BMC medicine},
  volume={15},
  number={1},
  pages={119},
  year={2017},
  publisher={Springer}
}

@article{hall2019moving,
  title={Moving beyond ‘safety’versus ‘autonomy’: a qualitative exploration of the ethics of using monitoring technologies in long-term dementia care},
  author={Hall, Alex and Brown Wilson, Christine and Stanmore, Emma and Todd, Chris},
  journal={BMC geriatrics},
  volume={19},
  number={1},
  pages={145},
  year={2019},
  publisher={Springer}
}

@article{denis2017web,
  title={Randomized trial comparing a web-mediated follow-up with routine surveillance in lung cancer patients},
  author={Denis, Fabrice and Lethrosne, Claire and Pourel, Nicolas and Molinier, Olivier and Pointreau, Yoann and Domont, Julien and Bourgeois, Hugues and Senellart, H{\'e}l{\`e}ne and Tr{\'e}moli{\`e}res, Pierre and Liz{\'e}e, Thibaut and others},
  journal={JNCI: Journal of the National Cancer Institute},
  volume={109},
  number={9},
  pages={djx029},
  year={2017},
  doi={10.1093/jnci/djx029},
  publisher={Oxford University Press}
}

@article{mooney2017automated,
  title={Automated home monitoring and management of patient-reported symptoms during chemotherapy: results of the symptom care at home {RCT}},
  author={Mooney, Kathi H and Beck, Susan L and Wong, Bob and Dunson, William and Wujcik, Debra and Whisenant, Margaret and Donaldson, Gary},
  journal={Cancer Medicine},
  volume={6},
  number={3},
  pages={537--546},
  year={2017},
  doi={10.1002/cam4.1002},
  publisher={Wiley Online Library}
}

@article{allsop2019paincheck,
  title={Multidisciplinary software design for the routine monitoring and assessment of pain in palliative care services: the development of {PainCheck}},
  author={Allsop, Matthew J and Johnson, Owen and Taylor, Sally and Hackett, Julia and Allen, Peter and Bennett, Michael I and Bewick, Bridgette M},
  journal={JCO Clinical Cancer Informatics},
  volume={3},
  pages={1--17},
  year={2019},
  doi={10.1200/CCI.18.00120},
  publisher={Wolters Kluwer Health}
}

@article{mao2024electronic,
  title={Electronic symptom monitoring for home-based palliative care: A systematic review},
  author={Mao, Suning and Liu, Liu and Miao, Cheng and Wang, Tianyi and Chen, Yue and Jiang, Zhishen and Hua, Chengge and Li, Chunjie and Cao, Yubin},
  journal={Palliative Medicine},
  volume={38},
  number={8},
  pages={790--805},
  year={2024},
  doi={10.1177/02692163241257578},
  publisher={SAGE Publications Sage UK: London, England}
}

@inproceedings{ahmadpour2023hci,
  title={How can {HCI} support end-of-life care? {C}ritical perspectives on sociotechnical imaginaries for palliative care},
  author={Ahmadpour, Naseem and Gough, Phillip and Lovell, Melanie and Austin, Philip and Poronnik, Philip and Zhang, Wendy Qi and Kay, Judy and Kummerfeld, Bob and Luckett, Tim and Brown, Martin and Phillips, Jane L and Agar, Meera},
  booktitle={Extended Abstracts of the 2023 CHI Conference on Human Factors in Computing Systems (CHI EA '23)},
  numpages={7},
  year={2023},
  doi={10.1145/3544549.3585768},
  publisher={Association for Computing Machinery},
  address={New York, NY, USA}
}

@article{yang2022confucian,
  title={Confucian familism and shared decision making in end-of-life care for patients with advanced cancers},
  author={Yang, Yuexi and Qu, Tingting and Yang, Jinyue and Ma, Ben and Leng, Anli},
  journal={International Journal of Environmental Research and Public Health},
  volume={19},
  number={16},
  pages={10071},
  year={2022},
  publisher={MDPI}
}

@incollection{dervin1992mind,
  title={From the Mind's Eye of the User: The Sense-Making Qualitative-Quantitative Methodology},
  author={Dervin, Brenda},
  booktitle={Qualitative Research in Information Management},
  editor={Glazier, Jack D and Powell, Ronald R},
  pages={61--84},
  year={1992},
  publisher={Libraries Unlimited},
  address={Englewood, CO}
}

@book{weick1995sensemaking,
  title={Sensemaking in Organizations},
  author={Weick, Karl E},
  year={1995},
  publisher={Sage Publications},
  address={Thousand Oaks, CA}
}

@article{weick2005organizing,
  title={Organizing and the Process of Sensemaking},
  author={Weick, Karl E and Sutcliffe, Kathleen M and Obstfeld, David},
  journal={Organization Science},
  volume={16},
  number={4},
  pages={409--421},
  year={2005},
  doi={10.1287/orsc.1050.0133}
}

@article{meng2026can,
  author = {Meng, Yibo and Zhan, Jinyuan and Lyu, Yao and Lu, Zhuoran and Zang, Chengxi},
  title = {``Can Computing Systems Eventually Disappear?'': A Long-Term Tracking Study on the Use of Diabetes Computing Systems by Patients with Diabetes},
  journal = {Proc. ACM Interact. Mob. Wearable Ubiquitous Technol.},
  year = {2026},
  month = dec,
  volume = {10},
  number = {4},
  articleno = {275},
  publisher = {Association for Computing Machinery},
  address = {New York, NY, USA},
  doi = {10.1145/3857983},
  url = {https://doi.org/10.1145/3857983}
}

@article{meng2026data,
  author = {Meng, Yibo and Shi, Shuoning and Liu, Bingyi and Chen, Hongkai},
  title = {When Data Becomes Judgment: Misaligned Interpretations and Accountability in Food Delivery Platforms},
  journal = {Proc. ACM Interact. Mob. Wearable Ubiquitous Technol.},
  year = {2026},
  month = dec,
  volume = {10},
  number = {4},
  articleno = {247},
  publisher = {Association for Computing Machinery},
  address = {New York, NY, USA},
  doi = {10.1145/3857972},
  url = {https://doi.org/10.1145/3857972}
}

@article{meng2026neglect,
  author = {Meng, Yibo and Liu, Ziqi and Wang, Lionel Z. and Zhang, Shuning and Ding, Xiaolan},
  title = {From Neglect to Resignation: Understanding the Evolution of Mobile Privacy},
  journal = {Proc. ACM Interact. Mob. Wearable Ubiquitous Technol.},
  year = {2026},
  month = dec,
  volume = {10},
  number = {4},
  articleno = {242},
  publisher = {Association for Computing Machinery},
  address = {New York, NY, USA},
  doi = {10.1145/3857977},
  url = {https://doi.org/10.1145/3857977}
}

@article{chen2026not,
  title={``Not Just Me and My To-Do List'': Understanding Challenges of Task Management for Adults with {ADHD} and the Need for {AI}-Augmented Social Scaffolds},
  author={Chen, Jingruo and Meng, Yibo and Nie, Kexin},
  journal={Proceedings of the ACM on Human-Computer Interaction},
  volume={10},
  number={6},
  pages={1--37},
  year={2026},
  publisher={ACM New York, NY}
}

@article{chen2026between,
  title={Between Knowledge and Care: A Mixed-Methods Evaluation of Generative {AI} for {T2DM} Self-Management from Patient and Physician Perspectives},
  author={Chen, Ruiqi and Meng, Yibo and Lu, Huidi and Ding, Xiaolan},
  journal={arXiv preprint arXiv:2607.03720},
  year={2026}
}

@article{meng2026tibetcpr,
  title={{TibetCPR}: A Multimodal Tactile Feedback System to Enhance Cardiopulmonary Resuscitation Training in High-Altitude Regions of {Tibet}},
  author={Meng, Yibo and Chen, Ruiqi and Liu, Zhiming and Ding, Xiaolan},
  journal={arXiv preprint arXiv:2606.07765},
  year={2026}
}

@article{meng2026living,
  title={Living Inside the Black Box: Behavioral Probing and Adaptation in Mandatory Wearable Sensing},
  author={Meng, Yibo and Liu, Bingyi and Chen, Ruiqi and Ding, Xiaolan and Ma, Shuai},
  journal={arXiv preprint arXiv:2607.09009},
  year={2026}
}

@article{meng2026does,
  title={Where Does the Human End? Creative Agency with Generative {AI} across Five Years of {Chinese} Digital Painting},
  author={Meng, Yibo and Chen, Ruiqi and Cao, Shuheng and Zhang, Weijia and Zang, Chengxi},
  journal={arXiv preprint arXiv:2609.09333},
  year={2026}
}

@article{meng2026engagement,
  title={Engagement Is Not Transfer: A Withdrawal Study of a Consumer Social Robot with Autistic Children at Home},
  author={Meng, Yibo and Fan, Guangrui and Liu, Bingyi and Sun, Yingfangzhong and Chen, Ruiqi and Mi, Haipeng},
  journal={arXiv preprint arXiv:2604.02642},
  year={2026}
}

@article{liu2025supporting,
  title={Supporting Our {AI} Overlords: Redesigning Data Systems to be Agent-First},
  author={Liu, Shu and Ponnapalli, Soujanya and Shankar, Shreya and Zeighami, Sepanta and Zhu, Alan and Agarwal, Shubham and Chen, Ruiqi and Suwito, Samion and Yuan, Shuo and Stoica, Ion and others},
  journal={arXiv preprint arXiv:2509.00997},
  year={2025}
}

@article{ma2026can,
  title={Can {AI} Agents Answer Your Data Questions? A Benchmark for Data Agents},
  author={Ma, Ruiying and Shankar, Shreya and Chen, Ruiqi and Lin, Yiming and Zeighami, Sepanta and Ghosh, Rajoshi and Gupta, Abhinav and Gupta, Anushrut and Gopal, Tanmai and Parameswaran, Aditya G},
  journal={arXiv preprint arXiv:2603.20576},
  year={2026}
}

@inproceedings{zhang2025slideaudit,
  title={{SlideAudit}: A Dataset and Taxonomy for Automated Evaluation of Presentation Slides},
  author={Zhang, Zhuohao and Chen, Ruiqi and Zhong, Mingyuan and Wobbrock, Jacob O},
  booktitle={Proceedings of the 38th Annual ACM Symposium on User Interface Software and Technology},
  pages={1--23},
  year={2025}
}
